\documentclass[manuscript,screen,review=false, nonacm]{acmart}
\usepackage{physics}

\AtBeginDocument{%
  \providecommand\BibTeX{{%
   \normalfont B\kern-0.5em{\scshape i\kern-0.25em b}\kern-0.8em\TeX}}}

\begin{document}

\title{Performance analysis of coreset selection for quantum implementation of K-Means clustering algorithm}


\author{Fanzhe Qu}
\affiliation{%
  \institution{School of Computing and Information Systems, The University of Melbourne}
  \city{Parkville}
  \state{Victoria}
  \country{Australia}}
  \postcode{3010}
  \email{fanzheq@student.unimelb.edu.au}
  
  \author{Sarah Monazam Erfani}
\affiliation{%
  \institution{School of Computing and Information Systems, The University of Melbourne}
  \city{Parkville}
  \state{Victoria}
  \country{Australia}}
  \postcode{3010}
  \email{sarah.erfani@unimelb.edu.au}
  
\author{Muhammad Usman}
\affiliation{%
  \institution{School of Physics, The University of Melbourne}
  \city{Parkville}
  \state{Victoria}
  \country{Australia}
  \postcode{3010}}
  \email{muhammad.usman@unimelb.edu.au}

\renewcommand{\shortauthors}{}

\begin{abstract}
Quantum computing is anticipated to offer immense computational capabilities which could provide efficient solutions to many data science problems. However, the current generation of quantum devices are small and noisy, which makes it difficult to process large data sets relevant for practical problems. Coreset selection aims to circumvent this problem by reducing the size of input data without compromising the accuracy. Recent work has shown that coreset selection can help to implement quantum K-Means clustering problem. However, the impact of coreset selection on the performance of quantum K-Means clustering has not been explored. In this work, we compare the relative performance of two coreset techniques (BFL16 and ONESHOT), and the size of coreset construction in each case, with respect to a variety of data sets and layout the advantages and limitations of coreset selection in implementing quantum algorithms. We also investigated the effect of depolarisation quantum noise and bit-flip error, and implemented the Quantum AutoEncoder technique for supressing the noise effect. Our work provides useful insights for future implementation of data science algorithms on near-term quantum devices where problem size has been reduced by coreset selection.

\end{abstract}
\maketitle

\section{Introduction} \label{intro}

Quantum computing is an emerging paradigm for data science problems, but a key fundamental question remains open: how can a quantum computer load a large classical scale data set and process it to find an answer? Today's quantum devices consist of only a few tens of qubits and their capabilities are limited due to noise. Contrarily real-world data science problems involve large volumes of data. For quantum algorithms to handle useful practical problems in near future, it is imperative to make meaningful size reduction of classical data sets so that they can be processed by the existing quantum devices without much loss of accuracy.  

In 2020, Harrow \cite{harrow2020small} proposed a quantum-classical hybrid method that uses coreset to break through the quantum limitation and offer quantum speedups for a variety of data science algorithms such as maximum likelihood estimation, Bayesian inference, and saddle-point optimization. A coreset is a small weighted set of data used to represent the large original data within an acceptable accuracy loss. In the same year, Tomesh \textit{et al.} \cite{tomesh2020coreset} proposed that the clustering of data can be transformed into a Hamiltonian optimization problem, with coreset as input, and solved with Quantum Approximate Optimization Algorithm  (QAOA) to realize quantum K-Means clustering. While Harrow \cite{harrow2020small} proposed that quantum speed-ups would be possible to achieve by implementing the coreset selection method and Tomesh \textit{et al.} \cite{tomesh2020coreset} studied the coreset construction technique for the K-Means clustering problem, the impact of coreset selection on results for data science problems by a quantum algorithm, such as on the performance of quantum K-Means clustering, remains an open question. For example, it is not well understood how a particular coreset selection scheme might effect the accuracy of a quantum algorithm?, or what is the impact of varying coreset size on the quantum algorithm implementation? The latter question also involves an interesting interplay between the improved algorithm accuracy due to larger coreset size and reduced fidelity due to stronger quantum noise present in the larger quantum circuit. Filling this gap could help researchers better understand the coreset selection, including its impact and its strengths and weaknesses, and ultimately advance this field. 

Our research aims to explore the impact of coreset selection on the  quantum K-Means clustering implementation and summarizes its advantages as well as the limitations of using coreset selection with a quantum algorithm. This paper explores two influencing factors of coreset selection on quantum K-Means clustering: (1) How the performance of quantum K-Means clustering changes when using different coreset construction algorithms; (2) How different coreset size interferes with quantum algorithm implementation. To answer the former question, we analysed the BFL16 coreset construction algorithm \cite{braverman2016new} and the ONESHOT coreset construction algorithm \cite{bachem2018one} which provide key insights into the influence of a particular coreset construction algorithm on quantum implementation. For (2), we varied coreset size from 5 to 10 and analysed its impact on the QAOA circuit performance. The evaluation method is to use the classical K-Means clustering as the benchmark to test the accuracy of the coreset on 6 different data sets, where the accuracy is defined as a direct benchmark against the classical K-Means clustering solution. Due to the limited quantum resources, we focused on 2-Means clustering problem as also the case in recently published work \cite{tomesh2020coreset}, but our methods can be readily generalised to the K-Means Clustering problem.  

Another important factor studied in our work is the impact of quantum noise. The depolarisation quantum noise and bit-flip errors will modify the quantum state, leading to erroneous results, interfering with the analysis, and eliminating the contribution of the coreset. In recent literature, there has been significant efforts to reduce the impact of noise by either implementing quantum error correction codes, or by implementing strategies at the quantum circuit level, such as Quantum AutoEncoders (QAEs), which inherits the idea of classical autoencoders, correcting quantum errors by learning fair parameters \cite{beer2019efficient}. The quantum error mitigation strategy tested in this paper is the QAEs, which compresses the input data to the bottleneck level to remove the noise.

Based on the analysed results, we find that both the BFL16 algorithm and the ONESHOT algorithm can achieve more than 95\% accuracy on all 6 data sets. The difference between these two algorithms is the coreset points selection preference. BFL16 tends to select the points of the coreset according to the distribution of the data set, while ONESHOT tends to learn the density of the data set. The impact of coreset size on accuracy is greater than that of the coreset construction algorithm. Experiments on various datasets show that generally increasing the size of the core set tends to reduce accuracy, but is data-dependent. The reason for this is that it is difficult to achieve a balance between coreset size and QAOA, although Nelder-Mead optimizer \cite{nelder1965simplex} is used to optimize QAOA to make it work efficiently, using more qubits makes the circuit larger and makes it harder for QAOA with Nelder-Mead optimizer to find the desired result. Larger circuits are also more susceptible to quantum noise. At first, the quantum noise is ignored to focus on the impact of coreset construction and selection. In the later part of the paper, noise is included and its impact is investigated on overall accuracy. A noise mitigation strategy, QAEs, is applied to mitigate the impact of quantum error and achieves positive results.
\\ \\ 
\noindent
The main contributions of our work sare as follows:
\begin{itemize}
\item We explore a relatively new area of research, coreset selection for quantum algorithm implementation, and present an analysis of the impact of coreset selection on the performance of quantum K-Means clustering.
\item We analyse the effect of the coreset size on the results, including the relationship between the coreset size and accuracy and the relationship between the coreset size and quantum algorithm.
\item We apply QAEs for the QAOA circuit and show that QAEs are effective strategy to mitigate the impact of quantum noise.
\end{itemize}

In the rest of the paper, Section \ref{review} reviews related papers and the background of the research. In Section \ref{method}, a detailed explanation of the used methods is provided. Performance analysis of coreset and QAEs will be in Section \ref{result}. Section \ref{conclusion} includes the conclusions of the paper and a discussion on future research directions.

\section{Related work AND Background}\label{review}

Performance analysis of coreset selection for quantum algorithms is a relatively new research field with only a few studies in the published literature to-date. In this section, we provide a summary of the related work on coreset selection and quantum denoising methods as well as the background of the research. 

\subsection{Coreset selection}
\subsubsection{Coreset construction}
Coreset is a weighted small data set used to represent the whole large data set. The naive approach to construct the coreset is to assign the same probability to each point in the data set and randomly select a small set as the coreset. However, the randomly selected coreset may not be able to effectively represent the original data because the sensitivity is not bounded. The sensitivity is the worst-case effect of each data point on the objective function \cite{braverman2016new}, and bounding the sensitivity helps to guarantee the result will always be approximately optimal. The bounding operation is realized by using $(\alpha,\beta) - approximation$ where $Dis(\mathcal{X},C) \leq \alpha \min Dis(\mathcal{X},C)$ and $|C| \leq \beta k$ while $\alpha$ and $\beta$ are small real number and $Dis$ represents the distance $Dis(\mathcal{X},\mathcal{C}) = |\mathcal{X} - \mathcal{C}|$. For K-Means clustering on coreset $C$ for data set $\mathcal{X}$, the objective function is $\sum_{x\in \mathcal{X}}\min |x - c_{closest}|$ where  $c_{closest}$ is the closest point in coreset $C$ for each data point $x$. In 2017, Bachem \textit{et al.} \cite{bachem2017practical} proposed a BFL16 coreset construction algorithm that measures the probability of each point being selected as coreset as the distance between each point and other points, and the $D^2$-sampling algorithm is used to limit the sensitivity. In this work, the BFL16 coreset has been applied to K-Means clustering, K-Median clustering, and Principal Component Analysis (PCA). The explanation for the steps it takes to construct the BFL16 coreset is:

\begin{enumerate}
\item Find $n$ K-Means++ cluster centers $B$ using $D^2$-sampling where $n$ is a hyper-parameter predefined based on the requirement and normally it will be a small integer value. The benefit of using K-Means++ cluster center instead of K-Means cluster center is that the K-Means++ clustering method can find more representative cluster centers and realize faster.
\item Set the probability of a certain point being selected as one of the elements in the coreset and the weight of this point based on the distance to $B$.
\item Construct the coreset $C$ of size $k$, where $k$ is the hyper-parameter indicates the number of elements in one coreset. Notice that the number of $k$ is the same as the number of qubits used when passing it to the quantum computer.
\end{enumerate}

Later in 2018, Bachem \textit{et al.} \cite{bachem2018one} proposed the ONESHOT coreset construction algorithm, using $D^2$-sampling to bound sensitivity. Different from the BFL16 algorithm, the $\Delta $ parameter is introduced to make the coreset more sensitive to the density of the data. The ONESHOT coreset has been proved to be suitable for dealing with multiple clustering problems at the same time. Similar to BFL16 algorithm, it also requires $n$ K-Means++ cluster centers $CC$ from $D^2$-sampling. The steps of the ONESHOT coreset are:

\begin{enumerate}
\item Preparing the sensitivity $S$ of each point. $S = \frac{\phi_{x}(CC) }{(2 * len)}$ where $\phi_{x}(CC)$ is obtained by dividing the distance of each point to its nearest $cc$ by the number of points in each cluster, and $len$ is the size of the data set. 
\item Define the hyper parameter $\Delta$that controls the gradient to $S$. It helps to make the coreset more close to a denser cluster. The sensitivity is bounded by the sum of sensitivities of all points $\{1,(1 + \Delta),...,P_{max}\}$, $P_{max}$ is a hyper-parameter that defines the upper bound.
\item The probability of each point to be selected as coreset is denoted as $P(x)$ . Equation of it is, $P(x) = \frac{S(x)}{\sum_{x'\in\mathcal{X}} S(x')}$. The weight of each coreset elements is $\frac{1}{len * m * P(x)}$ where $m$ is the size of coreset.

\end{enumerate}
\subsubsection{Coreset for quantum K-Means clustering}
Tomesh \textit{et al.} \cite{tomesh2020coreset} took BFL16 coreset as the input to the quantum algorithm, and used the quantum approximate optimization algorithm (QAOA) to obtain the state that gave the lowest Hamiltonian. The obtained state contains the information of splitting and clustering the coreset to form 2 cluster centroids. The metric to measure the method performance and accuracy loss is the cost $Cost = \sum_{c \in C, x \in \mathcal{X}} \|x-c\|^2$, which is the sum of squared distance from each point to the closest coreset point. In most cases, the cost of using the cluster centroids obtained from the quantum method was close to the cost of cluster centroids obtained directly using the 2-Means clustering model on the entire dataset. The accuracy loss is in a controllable range. Compared with Tomesh \textit{et al.} \cite{tomesh2020coreset}, our research is more focused on exploring the impact of coreset selection on the performance of quantum K-Means clustering, we define accuracy, shown in equation \ref{acc_equation}, as the criterion for evaluating performance, and the effects of different coreset construction algorithms, coreset size, QAOA performance and quantum noise on the performance of quantum K-Means clustering are analyzed.

\subsection{Quantum Approximate Optimization Algorithm}
QAOA is a quantum algorithm to find approximate solutions for optimization problems \cite{farhi2014quantum}. It starts with a trial state $\ket{\psi(\theta)}$ where $\theta$ is the gate parameter for any adjustable gate. It will then measure the expectation value of the system energy and use classical optimization algorithms to find a new $\ket{\theta}$ that gives lower system energy. By repeating these steps multiple times, it finally assigns a high probability to the state that gives the lowest system energy. In this research, the QAOA returns the state with the information on how to segment and cluster the input to form cluster centroids to minimize the objective function. The objective function for 2-Means clustering is shown in Equation \ref{objective function}, where $\mu$ is the set of cluster centroids. 
\begin{equation}
 \sum_{\mu_i \in \mu, x_i \in \mathcal{X}} min\|x_i-\mu_i\|^2
 \label{objective function}
\end{equation}
It has been proved that minimizing the objective function is equivalent to minimizing the squared distance from each point to its closest cluster centroids\cite{bishop2006pattern}, shown in Equation \ref{bishop}. The $S_{+1}$ is the cluster obtained by cluster centroid $\mu_{+1}$ and the $S_{-1}$ is the cluster obtained by cluster centroid $\mu_{-1}$.
\begin{equation}
 \sum_{i \in S_{-1}} \|x_i-\mu_{-1}\|^2 + \sum_{i \in S_{+1}} \|x_i-\mu_{+1}\|^2
 \label{bishop}
\end{equation}
The Equation \ref{bishop} can be transformed to maximising weighted inter-cluster distance, shown in Equation \ref{MAX-CUT equation}, where $W$ is the sum of weights in each cluster, the equation for $W$ is $W_{\pm1} = \sum_{i\in S_{\pm1}}w_i$, $w$ is the weight for each coreset element \cite{tomesh2020coreset}. It is worth mentioning that the $x$ here is no longer the point in the original data set but the elements in the coreset. Equation \ref{MAX-CUT equation} is also known as the MAX-CUT problem. In classical computation, this can be achieved by classical optimization algorithms such as Maximum Likelihood Estimation \cite{white1982maximum} or gradient descent. While in the quantum domain, the objective function of the MAX-CUT problem is transferred to Hamiltonian, and the QAOA is used to find the lowest Hamiltonian value. It will return a pair of symmetric states, telling us which partition maximizes the weighted inter-cluster distance.

\begin{equation}
 W_{+1}W_{-1}\lVert \mu_{+1} - \mu_{-1} \rVert^2
 \label{MAX-CUT equation}
\end{equation}

\subsection{Quantum denoising methods}
Quantum noise is a major obstacle to the performance of the quantum algorithm. To reduce the detrimental impact of errors, various quantum error mitigation and correction schemes have been proposed. Some quantum error correction algorithms like Shor's code \cite{calderbank1996good} and Steane's algorithm \cite{steane1996simple} work on the basic level of quantum computers, and some work on the circuit level such as Duplicated Circuit \cite{huggins2020virtual} and Quantum AutoEncoders \cite{achache2020denoising}. We will review the work that is on the circuit level.

\subsubsection{Duplicated Circuit}
Duplicated Circuit achieves the most accurate results by making $N$ copies of the circuit, and then take the average of the measurement results. The diagonalizing gates are assigned after the state preparation gates in each copied circuit. The matrix for diagonalizing gate is $\begin{bmatrix}
1 & 0 & 0 & 0\\
0 & \frac{\sqrt{2}}{2} & -\frac{\sqrt{2}}{2} & 0\\
0 & \frac{\sqrt{2}}{2} & \frac{\sqrt{2}}{2} & 0\\
0 & 0 & 0 & 1
\end{bmatrix}$. The limitation of Duplicated Circuit the state preparation gate required, such as $CNOT gate + Rotation Y gate + CNOT gate$. Since the QAOA circuit can not be converted to meet the formula of state preparation gate, the Duplicated Circuit can not be used as error correction strategy in this experiments.

\subsubsection{Quantum AutoEncoders}
In 2019, Beer \textit{et al.} \cite{beer2019efficient} proposed Quantum Neural Network (QNN), which inherits the idea of classical Artificial Neural Network (ANN). The hidden layers in QNN contain unitary operations $U$ instead of the activation function in ANN. The output state $\rho^{out}$ after feed-forward is in Equation \ref{feed forward}, where $\dagger$ is the transpose of the unitary operation, $tr$ is the trace of the matrix in linear algebra, and hidden layers $U^L = U_{nl}^LU_{nl-1}^L...U_1^L$, where $nl$ is the number of qubits in layer $L$. The parameters to be learned in QNN is the rotation angle in each rotation gate, so the backpropagation will update the rotation angles if the output state is different from the expectation.

\begin{equation}
  \ \rho^{out} = tr_{in,hidden}(U^{out}U^{L}...U^1(\rho^{in}\bigotimes |0...0>_{hidden,out}<0...0|)U^{1^{\dagger}...U^{L^\dagger}U^{out^{\dagger}}})
  \label{feed forward}
\end{equation}

Quantum AutoEncoders (QAEs) is an implementation of QNN for quantum denoising. It takes a pure quantum state as the expected output and a noisy state as input, performing feedforward and backpropagation to adjust parameters in the unitary operation. If the output after performing feed-forward is different from the expected input, backpropagation needs to be performed to allow the parameters in unitary operation to be updated. Through multiple feedforward and backpropagation, the unitary operation can almost perfectly learn how to map the corrupt state to the right state. Achache \textit{et al.} \cite{achache2020denoising} applied QAEs to correct the noisy Greenberger–Horne–Zeilinger (GHZ) state caused by bit-flips. From their work, the [2,1,2] QAEs can correct a noisy GHZ state back to a nearly pure state within 100 epochs.

\section{Methodology}\label{method}
\subsection{Research problem}
Our research aims to analyze the impact of coreset selection on quantum K-Means clustering performance from four perspectives: coreset construction algorithms, size of coreset, QAOA performance, and quantum noise. The performance is defined as: For a data set $D = \{d_1, d_2,...,d_n\}$, after applying classical 2-Means clustering, there will be two clusters $C_1 = \{d_1, d_2,...,d_k\}$ ($0 < k < n$) and $C_2 = \{d_{k+1},d_{k+2},...,d_n\}$, and $C_1$ and $C_2$ are taken as the benchmark. After applying quantum K-Means clustering, there are two quantum clusters $QC_1$ and $QC_2$, if $QC_1 = \{d_1, d_2,...,d_k\}$ and $QC_2 = \{d_{k+1},d_{k+2},...,d_n\}$, which means $QC_1 == C_1$ and $QC_2 == C_2$, then the performance of quantum K-Means clustering is prefect. The metric defined to measure performance is the accuracy shown in Section \ref{evaluation}. The performance of quantum K-Means clustering constructed by different coreset construction algorithms, different coreset sizes, and operating environments is measured to analyze the impact of these factors.

\subsection{Methods}
The performance of the BFL16 coreset construction algorithm \cite{braverman2016new} and ONESHOT algorithm \cite{bachem2018one} on 6 data sets are investigated to analyse the impact of the coreset construction algorithm. The distributions of coreset for 6 data sets are plotted to help understand the logic behind each algorithm and the difference between them. For both algorithms, 40 K-Means++ cluster centers constructed from $D^2$-sampling are taken as initial points. For ONESHOT algorithm, the hyper-parameter $\Delta = \frac{1}{log(len)}$ where $len$ is the length of data set and $P_{max} = 2$. 

In general, the larger the coreset size, the less accuracy is lost, and the better the representation of the original data. Experiments in \cite{tomesh2020coreset} prove this point. However, the results of this experiment showed that a larger coreset makes QAOA with Nelder–Mead optimizer difficult to do segmentation and clustering operations, and wrong QAOA result will lead to unsatisfactory final accuracy. The relationship diagram of coreset size and accuracy is provided for supporting analysis. These experiments are run on the noise-free quantum simulator to maximize the reliability of the results. 

\begin{figure}[h]
  \centering
  \includegraphics[width=0.7\linewidth]{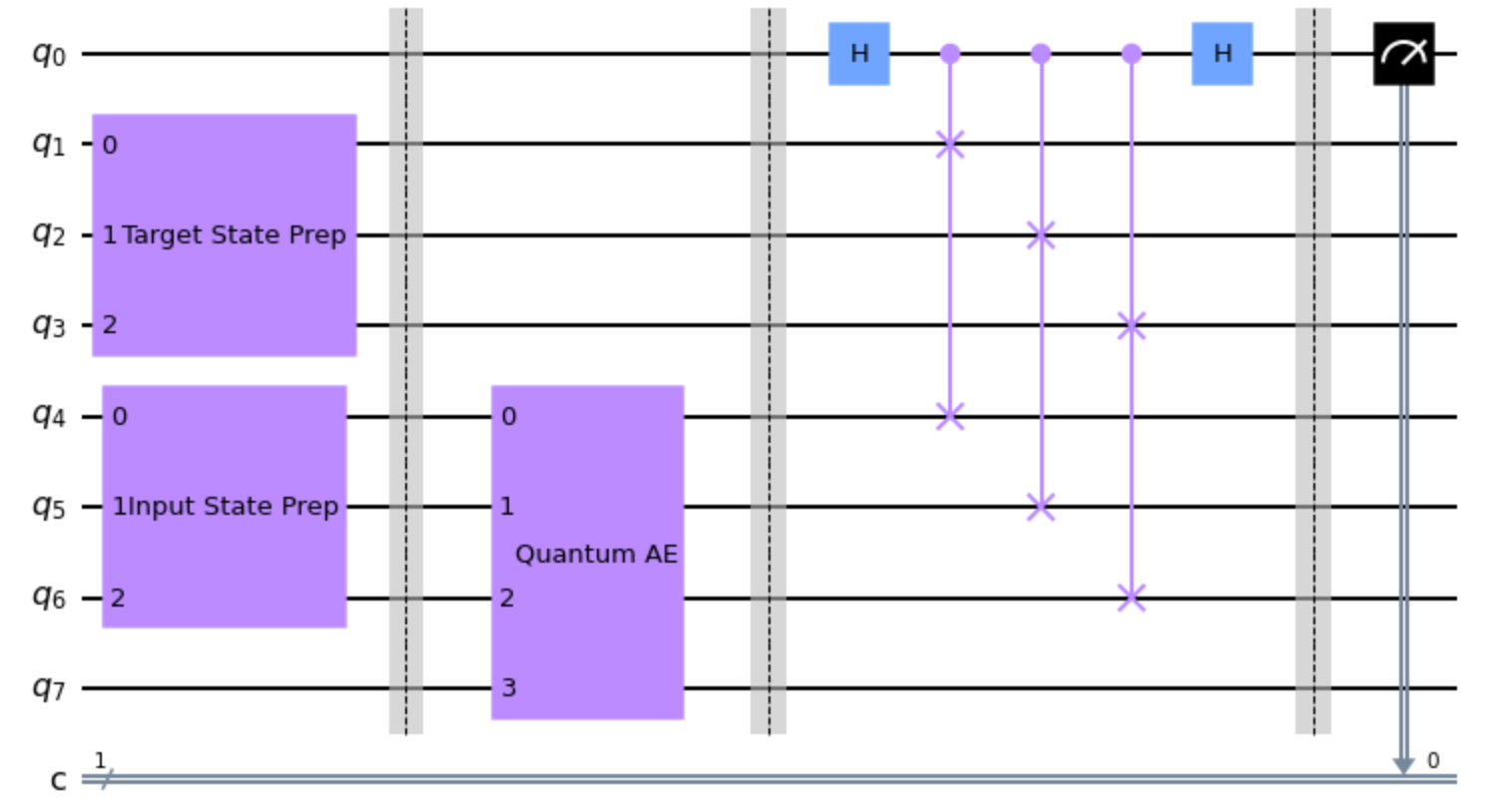}
  \caption{Circuits for QAEs}
  \label{QAEs_circuit}
\end{figure}

\begin{figure}[h]
  \centering
  \includegraphics[width=0.7\linewidth]{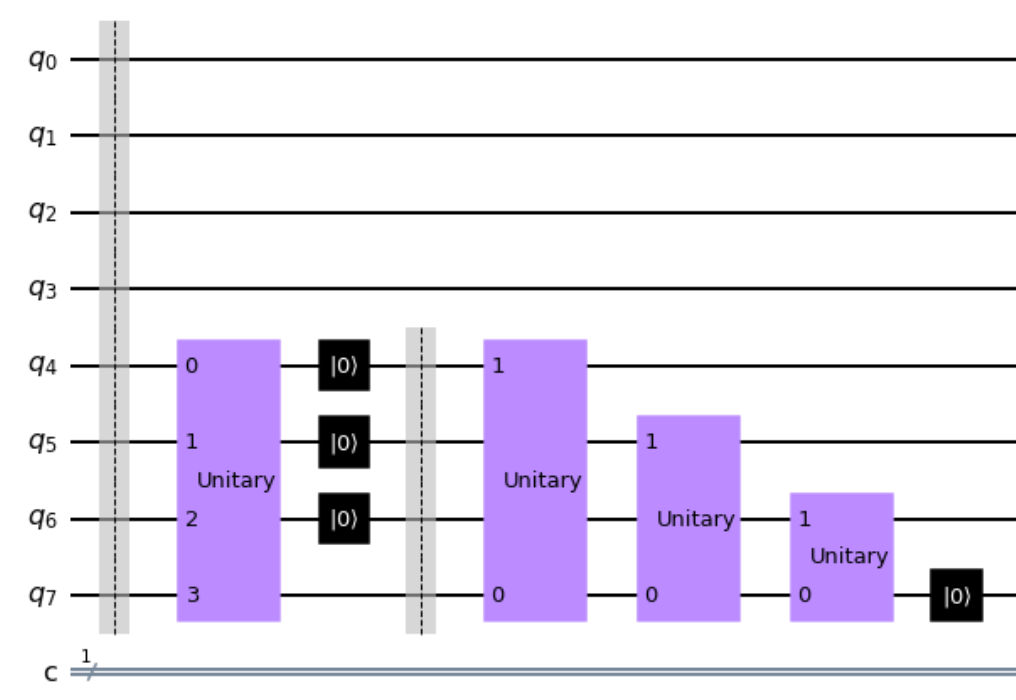}
  \caption{Decomposed circuits for QAEs}
  \label{Decomposed_QAEs}
\end{figure}

The depolarisation quantum noise and bit-flip error are introduced into the simulator from depolarizing noise model provided by IBM, the model allows the user to set the error rate and the type of gate where the error occurs, it will assign X gates to gates set by the user based on a set error rate. The Quantum AutoEncoders (QAEs) is applied to the QAOA circuit to get the collapsed state back to pure. Figure \ref{QAEs_circuit} shows the structure of [3,1,3] QAEs, that is, the input size is 3, the bottleneck layer size is 1, and the output layer size is 3. Target state prep and input state prep represent noise-free quantum state and corrupted quantum state respectively. The qubits occupied by QAEs is $1 + \textit{m}_1 + \textit{w}$, where $\textit{m}_1$ is the size of each layer and 1 is the first qubit used to measure state fidelity, $\textit{w}$ is the size of the QNN and its equation is $\textit{w} = \mathop{max}\limits_{1 \le i \leq (len - 1)}(\textit {m}_i + \textit{m}_{i+1})$. For [3,1,3] structure, $\textit{m}_1$ is 3 because the input size is 3, $\textit{w}$ is 4, so there are totally 8 qubits used. The H gate and CCX gate between the second and third barriers are used to compare the inconsistency of the input and target states and reflect it through the first qubit. Figure \ref{Decomposed_QAEs} shows the decomposition circuit of QAEs. The unitary operation is $U = e^{ik}$ where $k = \sum_{\sigma \in P^{\bigotimes n}}K_ \sigma \sigma$, where $n$ is the number of qubits used in the circuit, $P \in \{I, Y, X, Z\}$, an example of $P^{\bigotimes 3} \in\{ III, IXX,IXY...\}$, $K$ are the coefficients of the unitary operation vector. The training process is to train coefficient $K$that makes the input state close to the pure state as much as possible.

\section{RESULTS AND DISCUSSION} \label{result}
The objective of the experiment is to demonstrate to what extent quantum K-Means clustering is affected by our proposed influencing factors.

\subsection{Evaluation} \label{evaluation}
The performance of quantum K-Means clustering is analysed by setting accuracy as a metric to measure the degree of influence of each factor on quantum K-Means clustering. Accuracy represents the use of quantum K-Means clustering, how many points have been assigned into the same cluster as classical K-Means clustering. The equation of accuracy is:
\begin{equation}
  \label{acc_equation}
  \ Accuracy=\dfrac{\text{number of correctly assigned points}}{\text{size of data set}}
\end{equation}
 Table \ref{Datasets} describes 6 data sets used in experiments. The selection of data sets is based on their size and dimension, ranging from the Physical exercise data set with a small amount of data and low dimension to the Epilepsy dataset with a large amount of data and a high dimension.

\begin{table*}
  \caption{Data sets}
  \label{Datasets}
  \begin{tabular}{ccl}
    \toprule
    Data sets & Number of points/Dimension &Description\\
    \midrule
    \texttt{Physical exercise} & 20/3& Physical exercise Linnerud data set.\cite{sklearn_api}\\
    \texttt{Iris}& 150/4 & Iris data set contains 150 instances for 3 different type of iris plant.\cite{fisher1936use}\\
    \texttt{Wine}& 178/13 & Wine data set shows 13 components in three types of wine. \cite{hettichc1998blake}\\ 
    \texttt{Boston house price} & 506/13& House prices in the Boston area and possible influencing factors. \cite{harrison1978hedonic}\\
    \texttt{Breast cancer} & 569/30 &This dataset records information about breast cancer patients.\cite{hettichc1998blake}\\
    \texttt{Epilepsy} & 11.5k/179& Data set used to recognize epileptic seizure.\cite{andrzejak2001indications}\\
    \bottomrule
  \end{tabular}
\end{table*}

\subsection{Coreset construction algorithm} \label{cca}

Table \ref{coreset algorithm performance} presents the accuracy and standard deviation of the coreset obtained by BFL16 and ONESHOT algorithms on the six data sets. "Accuracy - quantum" in the table refers to the accuracy of quantum k-means clustering obtained by coreset and QAOA, and "Accuracy-classical" refers to the accuracy of K-Means clustering obtained by coreset and weighted average on a classical computer. The accuracy in the table is the average accuracy after ten experiments, not the highest one. Accuracy on both quantum and classical is obtained from the coreset size of 5 for 6 data sets. Regardless of the coreset construction algorithm used, using QAOA, or using a weighted average to do the segmentation and clustering operations on a classical computer, an accuracy of more than 90\% can be achieved. 

\begin{figure}[h]
 \begin{minipage}{0.32\linewidth}
 \vspace{3pt}
 \centerline{\includegraphics[width=\textwidth]{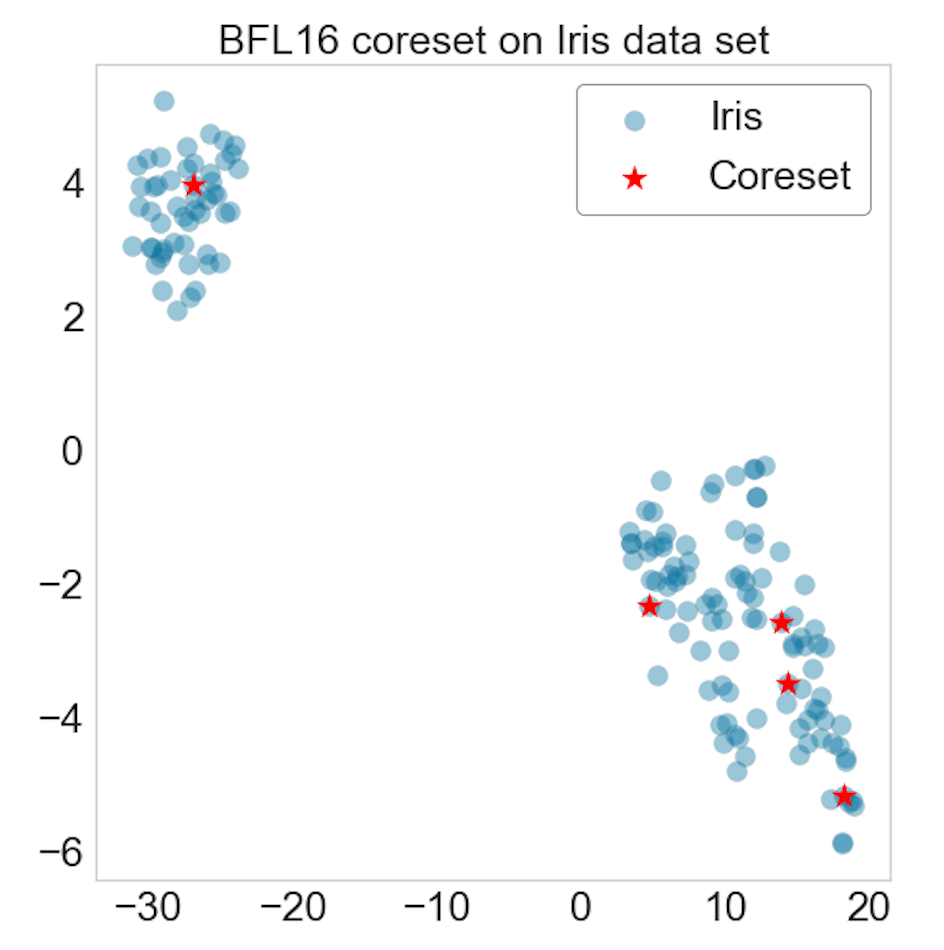}}
 \vspace{3pt}
 \centerline{\includegraphics[width=\textwidth]{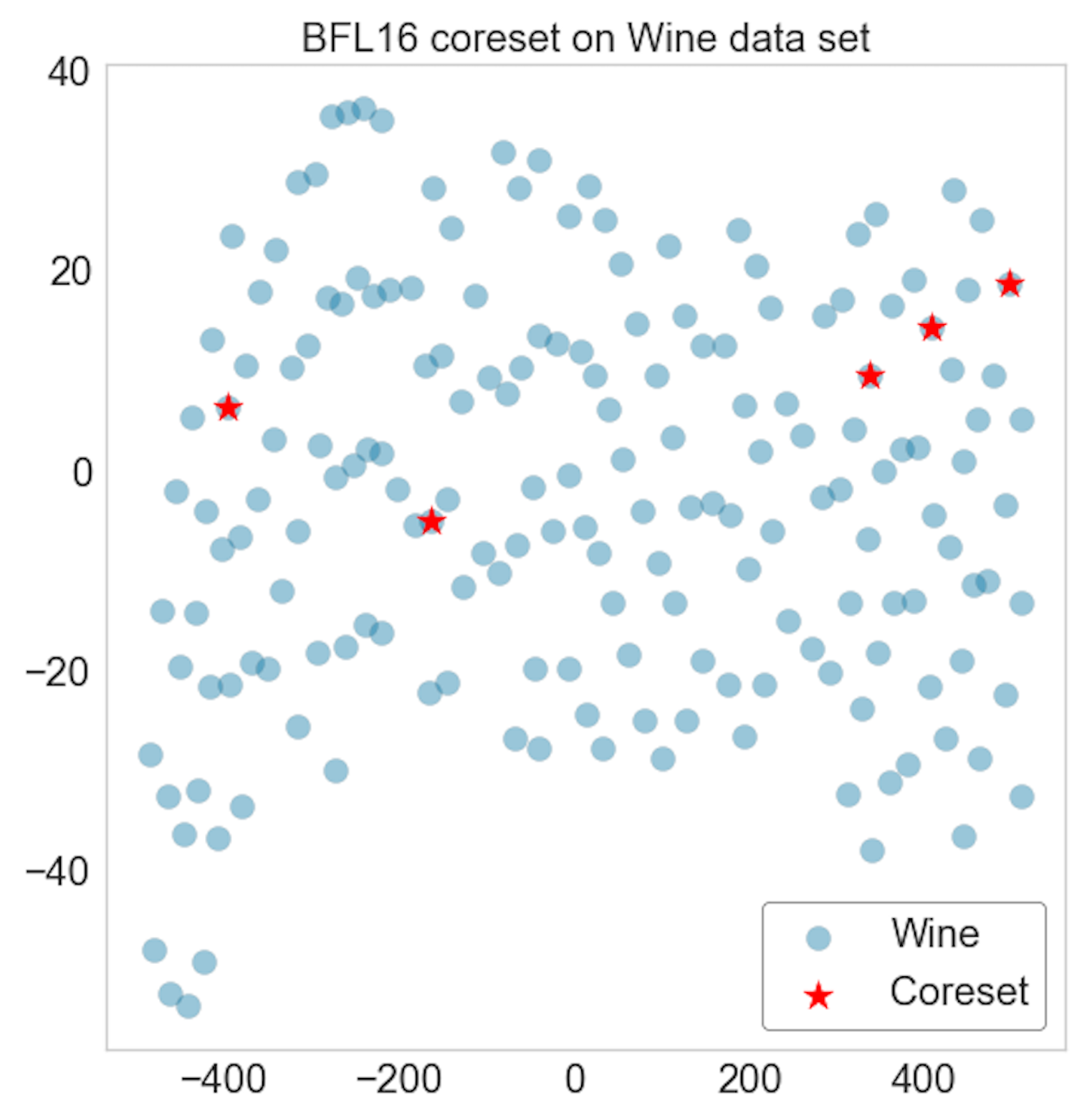}}
 \vspace{3pt}
 \end{minipage}
 \begin{minipage}{0.32\linewidth}
 \vspace{3pt}
 \centerline{\includegraphics[width=\textwidth]{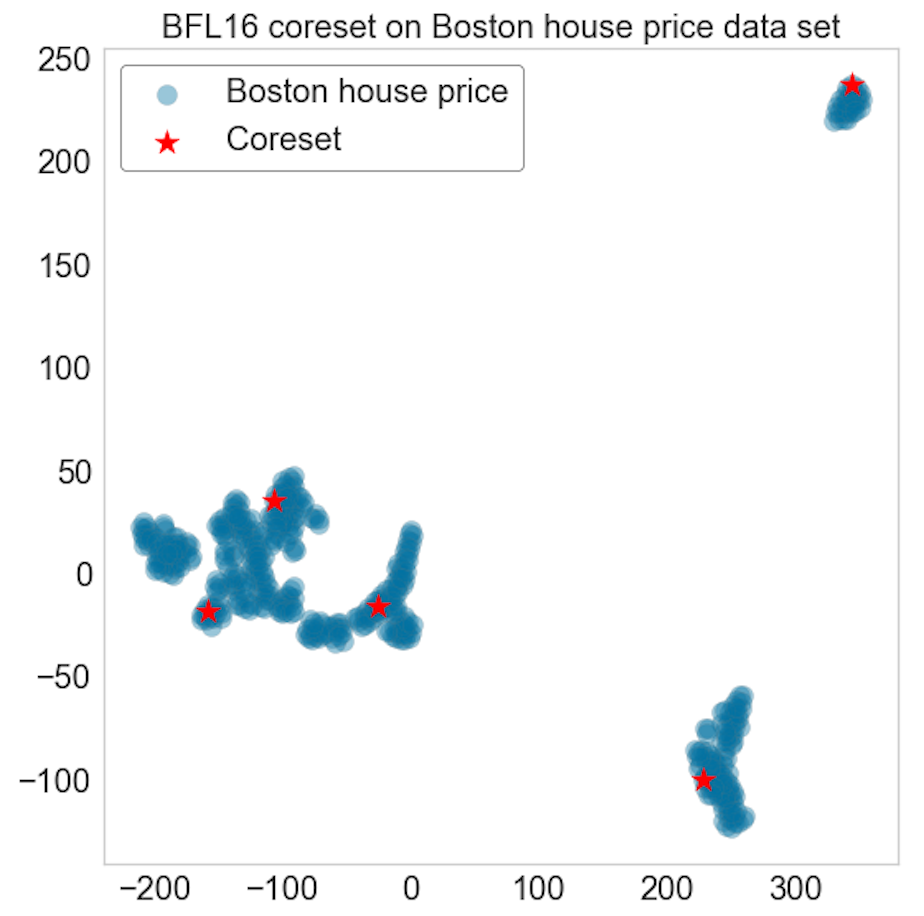}}
 \vspace{3pt}
 \centerline{\includegraphics[width=\textwidth]{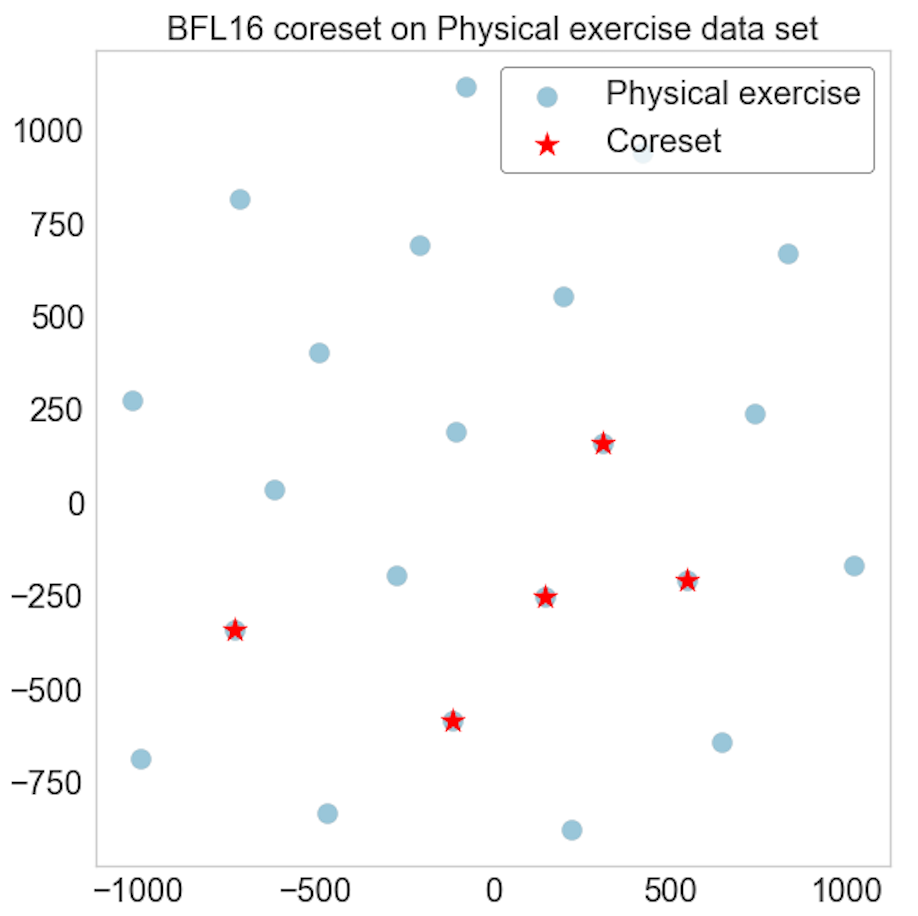}}
 \vspace{3pt}
 \end{minipage}
 \begin{minipage}{0.32\linewidth}
 \vspace{3pt}
 \centerline{\includegraphics[width=\textwidth]{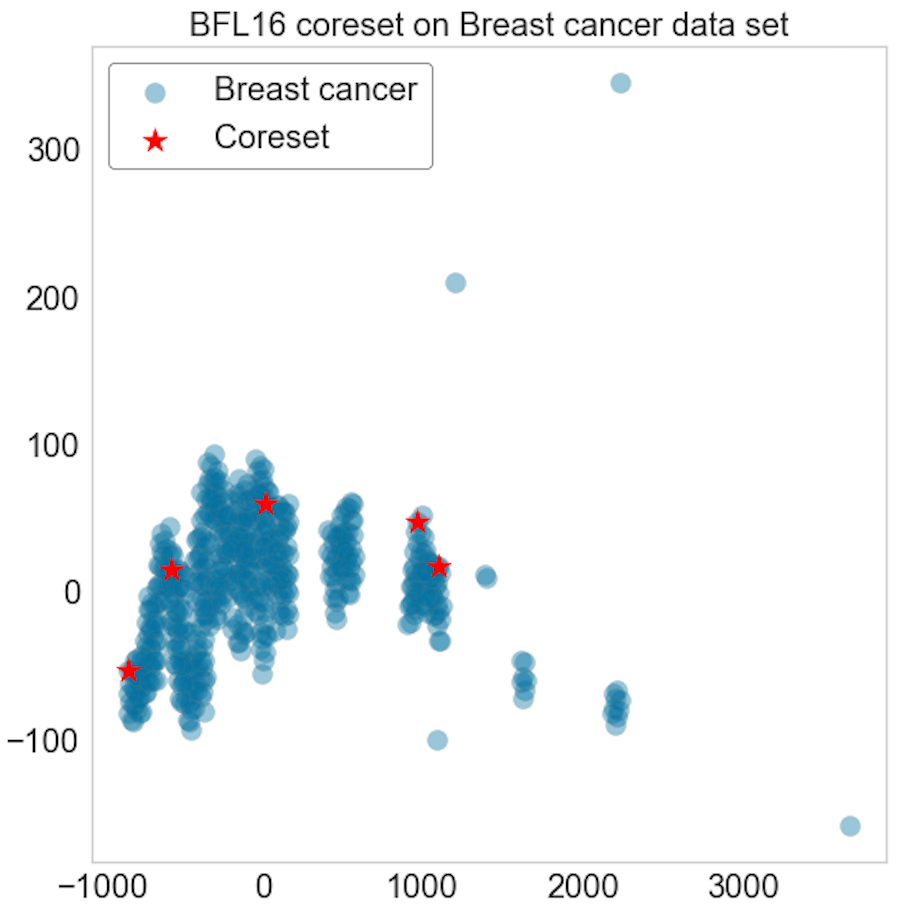}}
 \vspace{3pt}
 \centerline{\includegraphics[width=\textwidth]{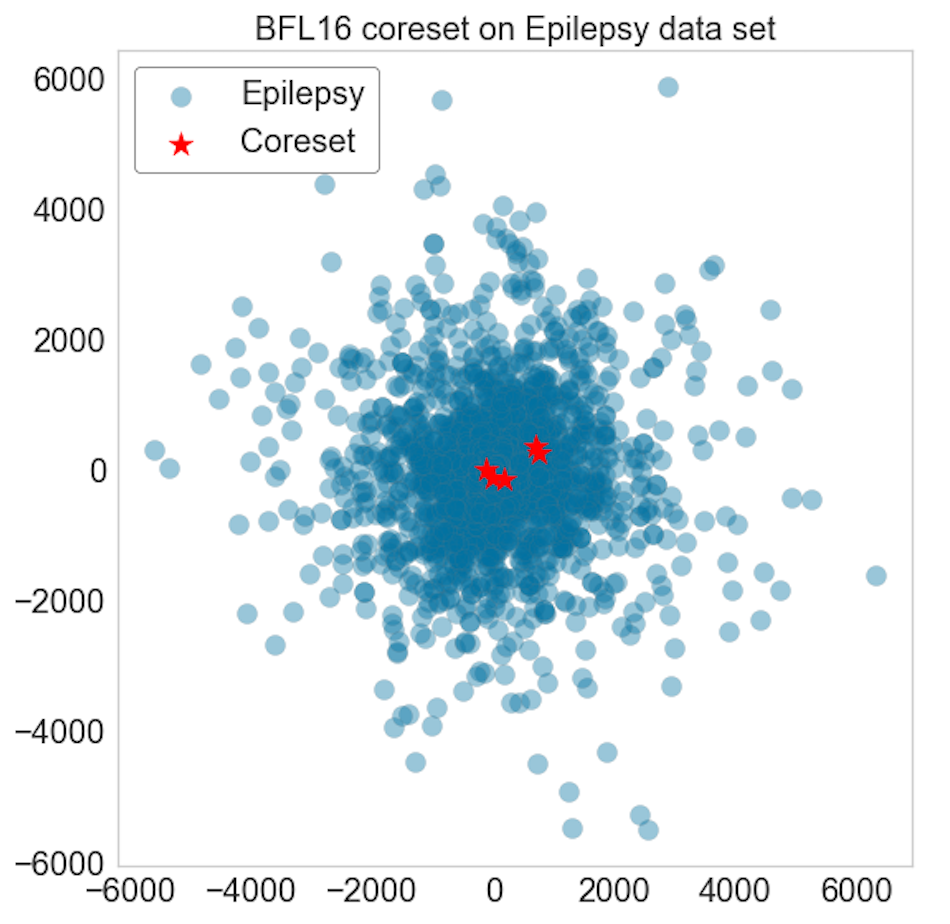}}
 \vspace{3pt}
 \end{minipage}
 \caption{Implementation of BFL16 coreset construction algorithm on 6 data sets}
 \label{BFL16}
\end{figure}

\begin{figure}[h]
 \begin{minipage}{0.32\linewidth}
 \vspace{3pt}
 \centerline{\includegraphics[width=\textwidth]{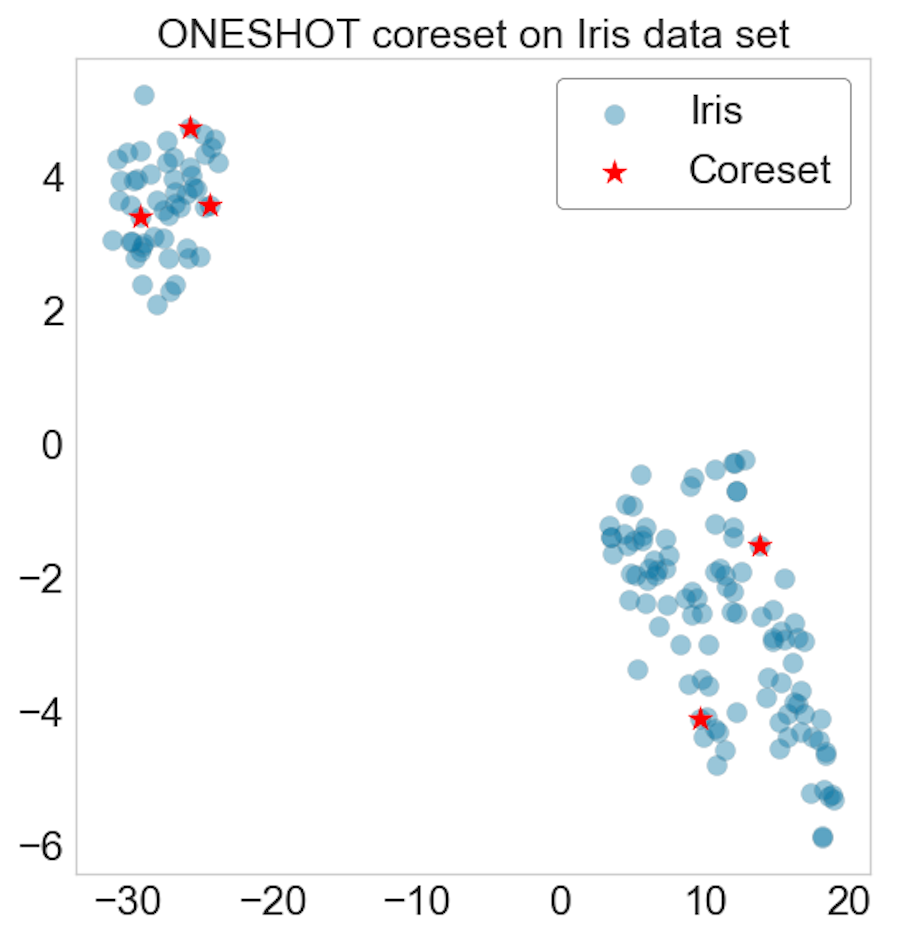}}
 \vspace{3pt}
 \centerline{\includegraphics[width=\textwidth]{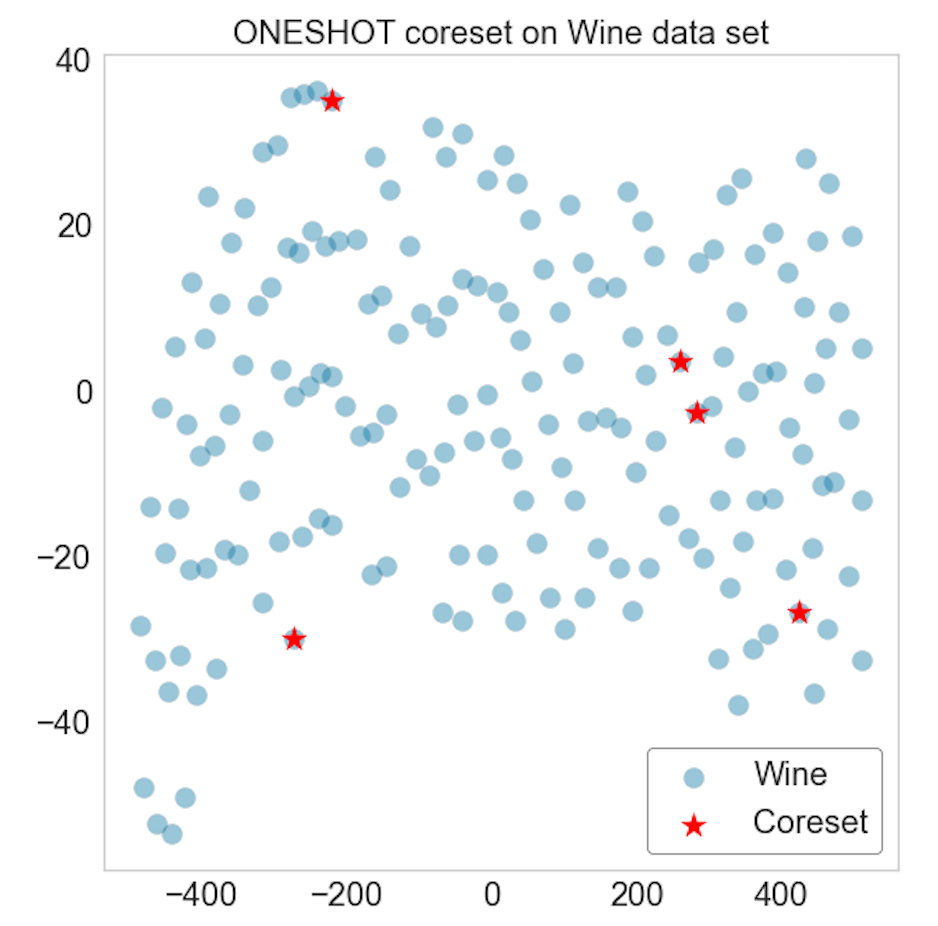}}
 \vspace{3pt}
 \end{minipage}
 \begin{minipage}{0.32\linewidth}
 \vspace{3pt}
 \centerline{\includegraphics[width=\textwidth]{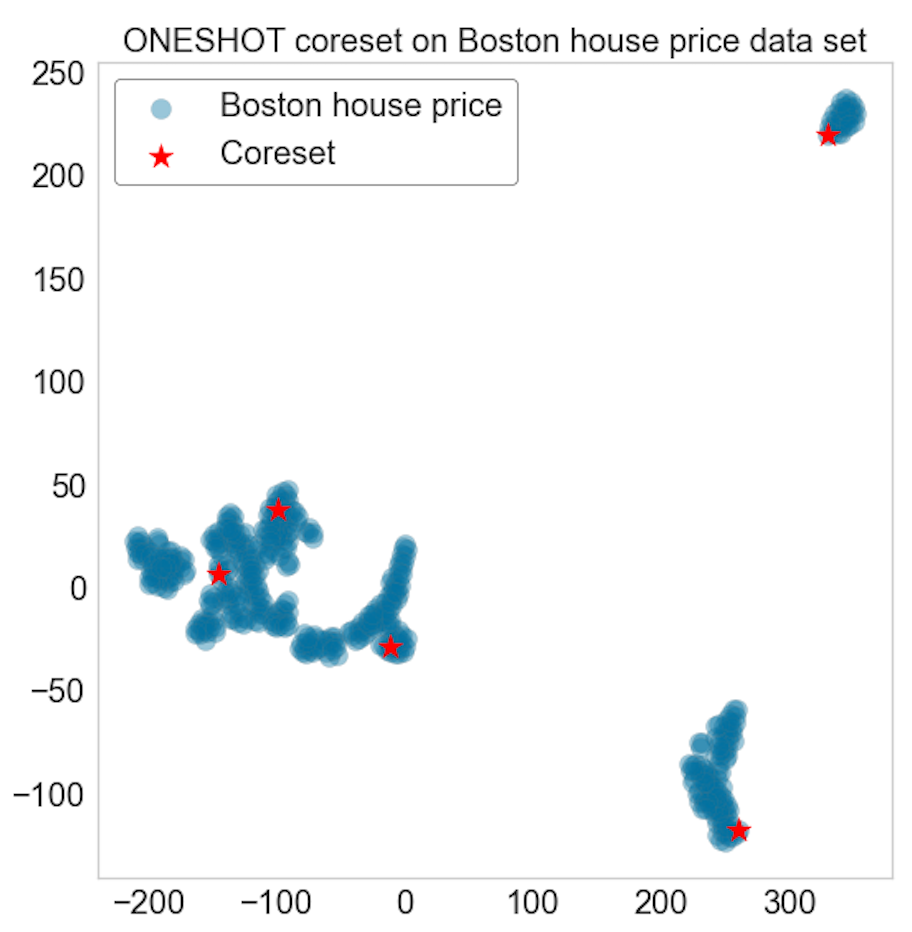}}
 \vspace{3pt}
 \centerline{\includegraphics[width=\textwidth]{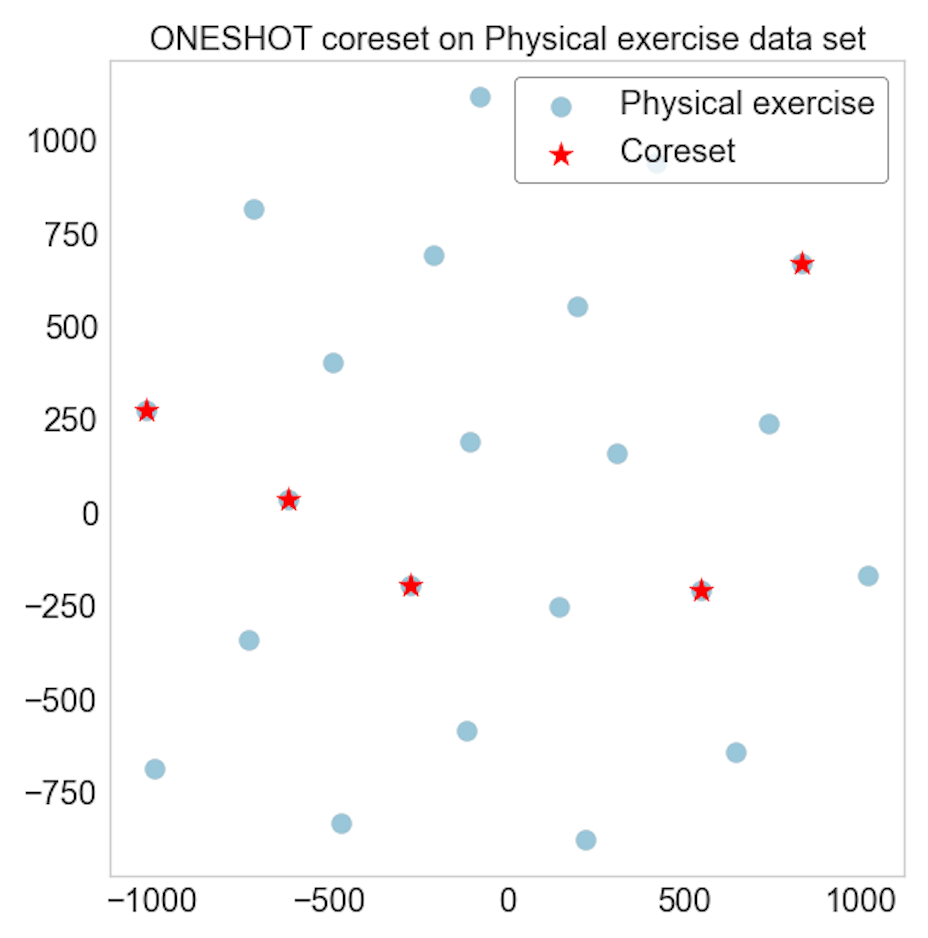}}
 \vspace{3pt}
 \end{minipage}
 \begin{minipage}{0.32\linewidth}
 \vspace{3pt}
 \centerline{\includegraphics[width=\textwidth]{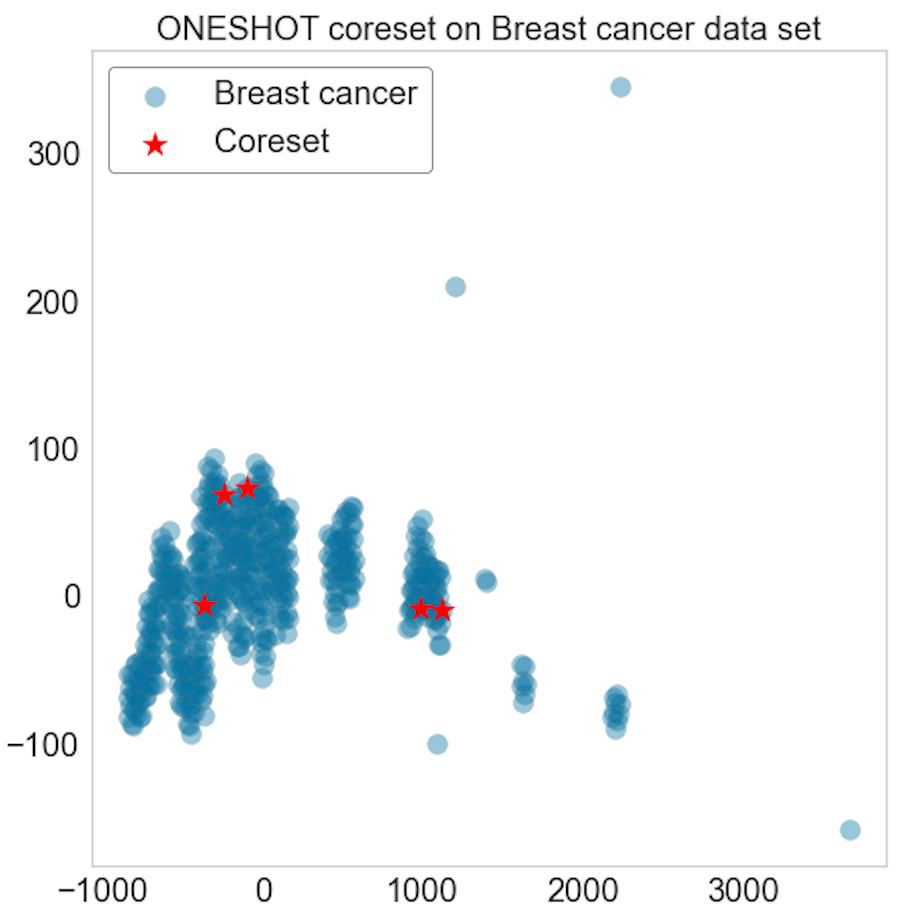}}
 \vspace{3pt}
 \centerline{\includegraphics[width=\textwidth]{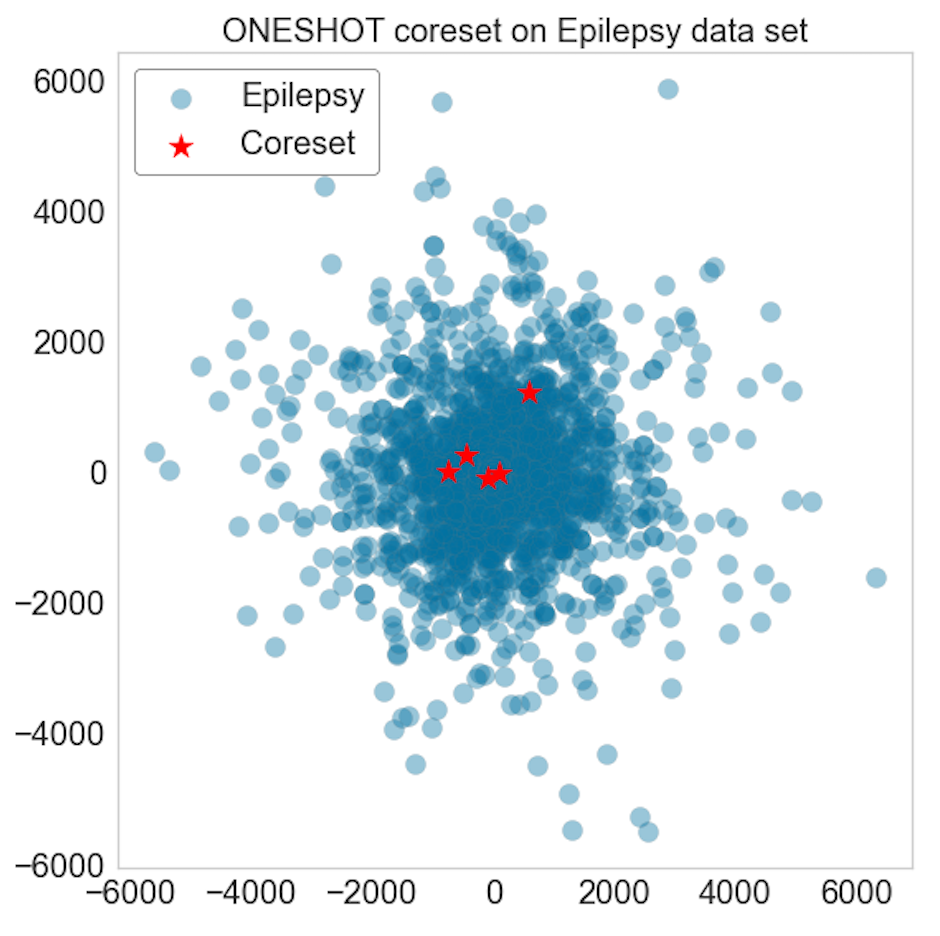}}
 \vspace{3pt}
 \end{minipage}
 \caption{Implementation of ONESHOT coreset construction algorithm on 6 data sets}
 \label{ONESHOT}
\end{figure}

The data in "Accuracy-classical" is not affected by quantum algorithms and can visualize the impact of coreset construction. For all 6 data sets, both the BFL16 and ONESHOT algorithms achieve at least 95\% accuracy, which means the coreset from two algorithms represents the original large data well. The difference in accuracy on 6 data sets mainly comes from the distribution of the data set itself. For the robustly clustered data sets such as Iris data set, Boston housing price data set, and Breast cancer data set, both BFL16 and ONESHOT coreset can achieve good accuracy. The impact of dimension and number of points in the data set is limited because data sets like physical exercise and wine are not as large and high-dimension as the Boston house price and breast cancer data set, and the accuracy of the coreset construction algorithm on them is low mainly because of their data distribution. The distribution of the data set and the distribution of the coreset constructed by BFL16 and ONESHOT are presented in Figure \ref{BFL16} and Figure \ref{ONESHOT} through T-distributed Stochastic Neighbor Embedding (TSNE) and PCA. However, if using a coreset on a data set with a large amount of data and high dimensions like the Epilepsy data set, the performance of the coreset will still be affected.

The BFL16 algorithm tends to follow the distribution of the data and assign more points to the larger cluster; while the coreset of the ONESHOT algorithm follows the density of the data and assigns most points to the cluster with high density. For example, the Iris data set has two clusters, and the cluster in the lower right corner is relatively large and wide. BFL16 algorithm places 4 of the 5 points in that area, while the ONESHOT algorithm only gives two. ONESHOT algorithm gives three points to the denser cluster in the upper left corner, while over there, the BFL16 algorithm gives only one. The difference comes from the parameter. $\Delta$ in the ONESHOT algorithm makes adjustments to the coreset and makes the coreset sensitive to the density of data, while BFL16 determines the coreset by calculating the distance between points so that it prefers to learn the distribution of data.

\begin{table}[]
  \caption{Performance of coreset construction algorithms with coreset size 5 on 6 data sets (accuracy $\pm$ standard deviation)}
  \label{coreset algorithm performance}
  \begin{tabular}{|c|cc|cc|}
  \hline
  Data sets              & \multicolumn{2}{c|}{BFL16}                                                        & \multicolumn{2}{c|}{ONESHOT}                                                      \\    \hline
  \multicolumn{1}{|l|}{} & \multicolumn{1}{l|}{Accuracy - quantum} & \multicolumn{1}{l|}{Accuracy - classical} & \multicolumn{1}{l|}{Accuracy - quantum} & \multicolumn{1}{l|}{Accuracy - classical} \\ \hline
  Physical exercise      & \multicolumn{1}{c|}{94\%  $\pm$ 0.08}         & 97\% $\pm$ 0.03                               & \multicolumn{1}{c|}{95\% $\pm$ 0.03}          & 99\% $\pm$ 0.02                               \\ \hline
  Iris                   & \multicolumn{1}{c|}{92\% $\pm$ 0.1}           & 99.4\% $\pm$ 0                                & \multicolumn{1}{c|}{94\% $\pm$ 0.06}          & 99.2\% $\pm$ 0                                \\ \hline
  Wine                   & \multicolumn{1}{c|}{91.8\% $\pm$ 0.07}        & 97.6\% $\pm$ 0.01                             & \multicolumn{1}{c|}{90\% $\pm$ 0.08}          & 97.6\% $\pm$ 0.01                               \\ \hline
  Boston house price     & \multicolumn{1}{c|}{92.3\% $\pm$ 0.08}        & 99.9\% $\pm$ 0                                & \multicolumn{1}{c|}{94\% $\pm$ 0.05}          & 100\%                                   \\ \hline
  Breast cancer          & \multicolumn{1}{c|}{92.1\% 0.06}        & 97.6\% $\pm$ 0.02                             & \multicolumn{1}{c|}{91.7\% $\pm$ 0.06}        & 96.8\% $\pm$ 0.02                                  \\ \hline
  Epilepsy               & \multicolumn{1}{c|}{91.6\% $\pm$ 0.12}        & 95.9\% $\pm$ 0                                & \multicolumn{1}{c|}{92.4\% $\pm$ 0.05}        & 95.8\% $\pm$ 0                                \\ \hline
  \end{tabular}
 \end{table}

\subsection{Quantum Approximate Optimization Algorithm}

\begin{figure}[h]
  \centering
  \includegraphics[width=0.5\textwidth]{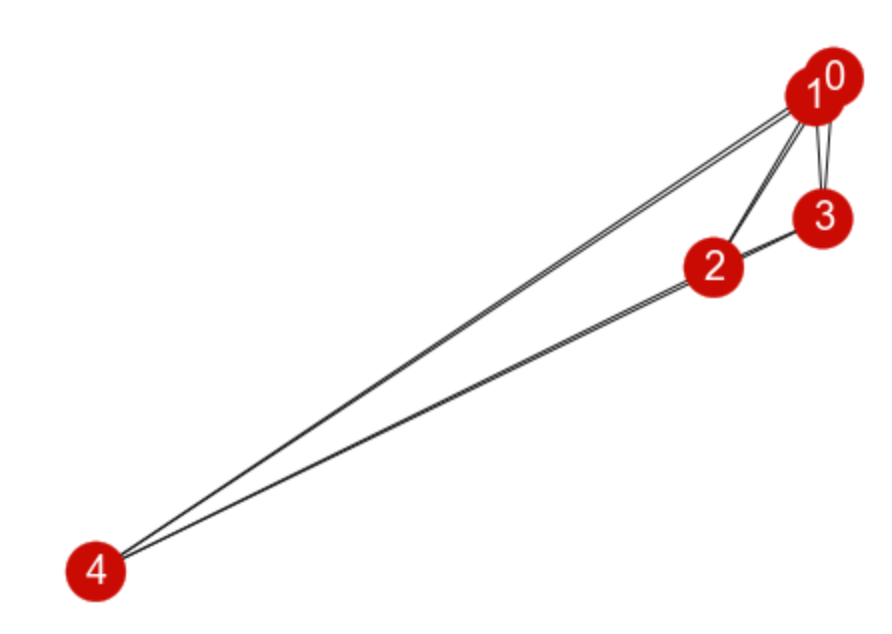}
  \caption{Weighted complete graph for Epilepsy coreset of size 5}
  \Description{Weighted complete graph for Epilepsy coreset of size 5.}
  \label{complete_graph}
\end{figure}

\begin{figure}[h]
  \centering
  \includegraphics[width=\linewidth]{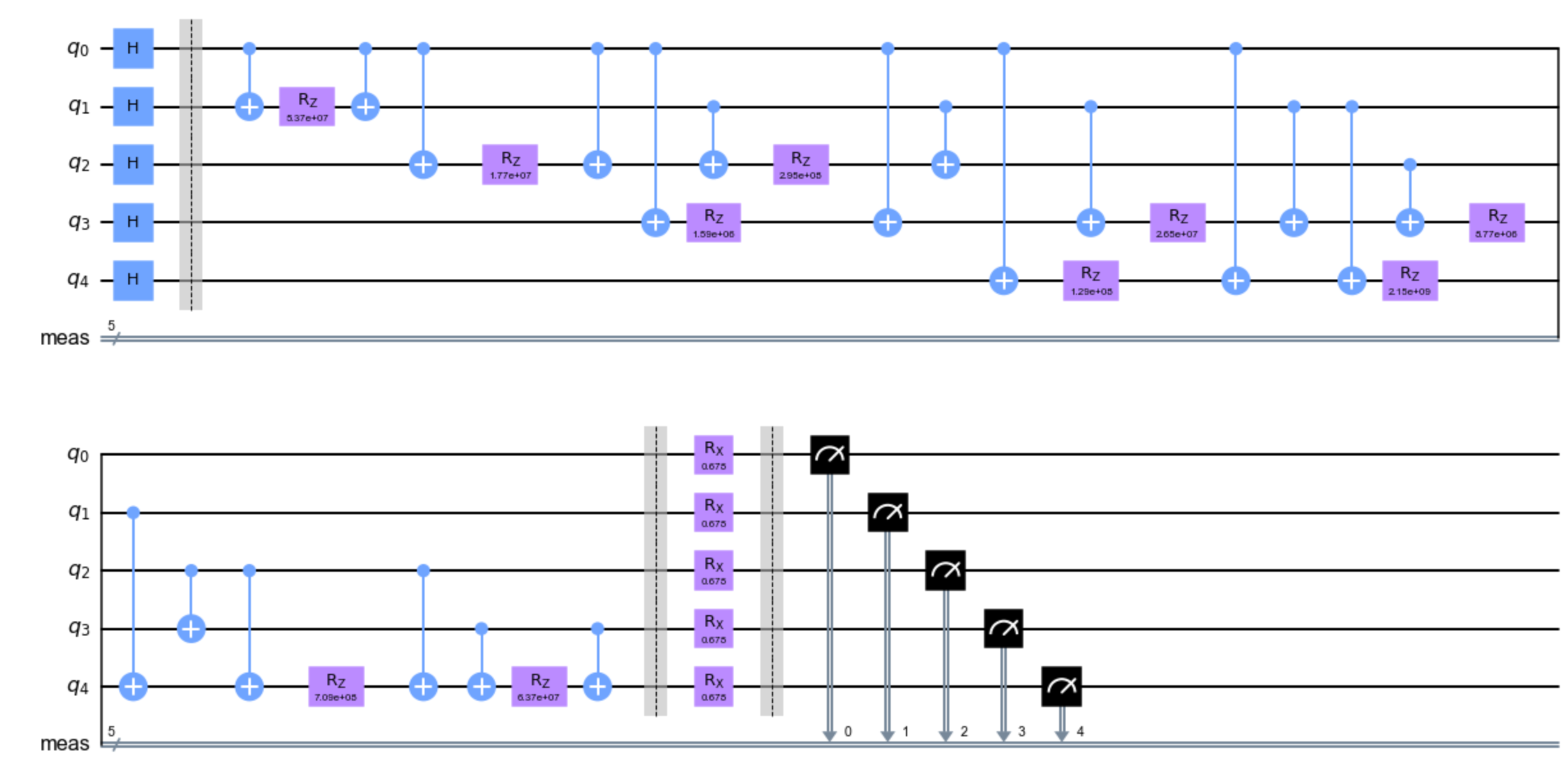}
  \caption{QAOA circuit for Epilepsy coreset of size 5}
  \Description{QAOA circuit for Epilepsy coreset of size 5.}
  \label{QAOA_circuit}
\end{figure}

\begin{figure}[h]
  \centering
  \includegraphics[width=0.8\textwidth]{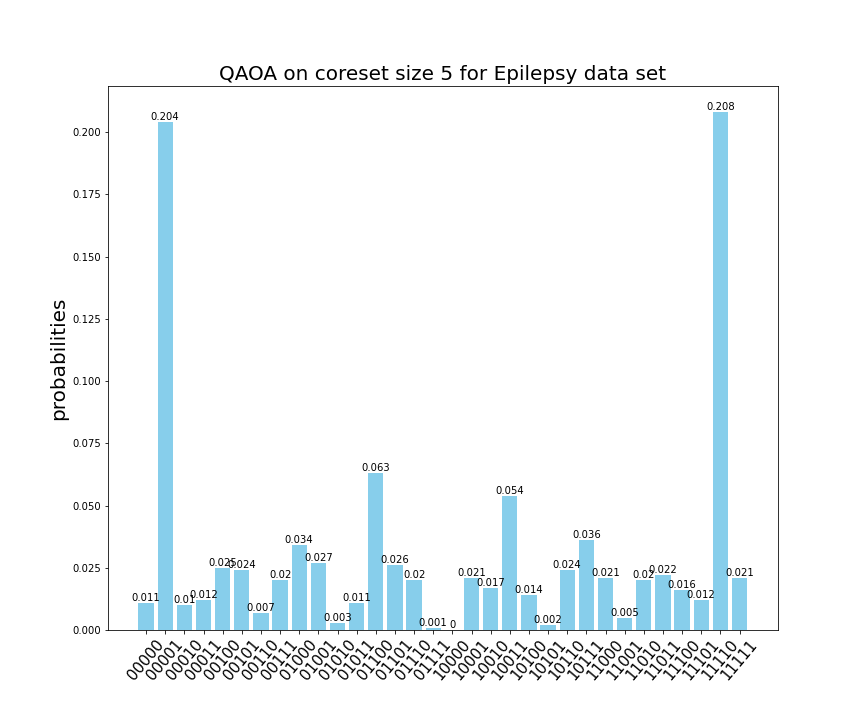}
  \caption{QAOA result for Epilepsy coreset of size 5}
  \Description{QAOA result for Epilepsy coreset of size 5.}
  \label{QAOA_result}
\end{figure}

In the classical K-Means clustering algorithm, the cluster centroids are formed by calculating the distance between points to form $k$ clusters and then calculating the weighted average in each cluster. The quantum K-Means clustering constructs cluster centroids through QAOA. It converts the weighted coreset into a weighted complete graph and solves it as a weighted MAX-CUT problem. Figure \ref{complete_graph} shows a weighted complete graph obtained using a coreset of size 5 on the Epilepsy data set, indexed from 0 to 4, and drawn by the nextworkx library. The Hamiltonian for the weighted coreset is, $Ham= -9887.5ZZIII + 5135.7ZIZII -1855.3ZIIZI + -15242.2ZIIIZ -20656.1IZZII + 7462.0IZIZI + 61305.0IZIIZ -3875.9IIZZI -31842.8IIZIZ + 11503.2IIIZZ $, and then QAOA is applied to obtain the maximum Hamiltonian. The QAOA circuit is shown in Figure \ref{QAOA_circuit}. The $H$ gate is used to entangle the qubits so that the $CNOT$ gates can make a controlled change, the last 5 gates are used to measure the results. Figure \ref{QAOA_result} shows that there are two quantum states that have the highest probability, $\ket{11110}$ and $\ket{00001}$, which are a pair of symmetric states, $i.e$. the coreset points with indices 0 to 3 are in one cluster, and index 4 is in the other cluster. 

\begin{figure}[h]
 \begin{minipage}{0.32\linewidth}
 \vspace{3pt}
 \centerline{\includegraphics[width=\textwidth]{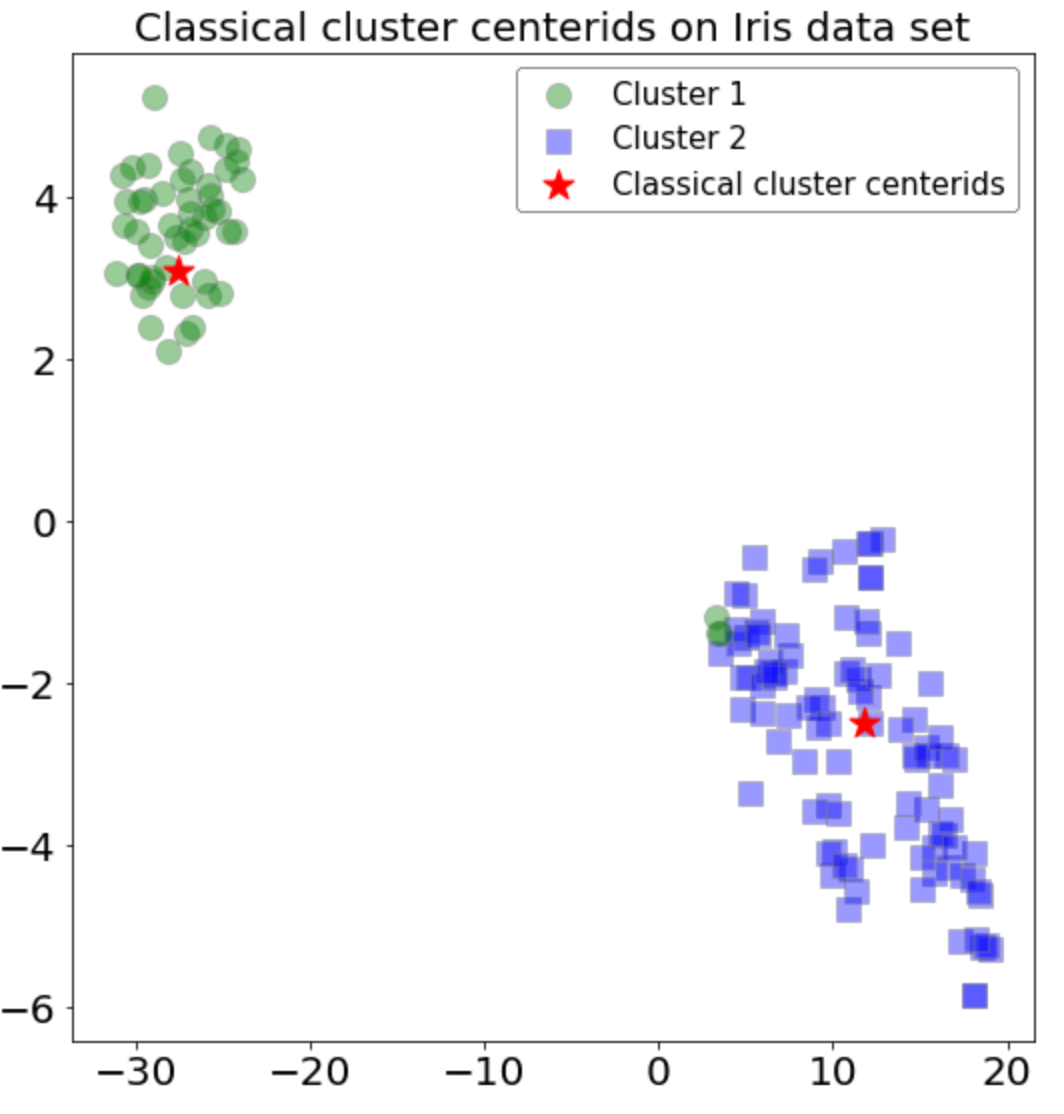}}
 \vspace{3pt}
 \end{minipage}
 \begin{minipage}{0.32\linewidth}
 \vspace{3pt}
 \centerline{\includegraphics[width=\textwidth]{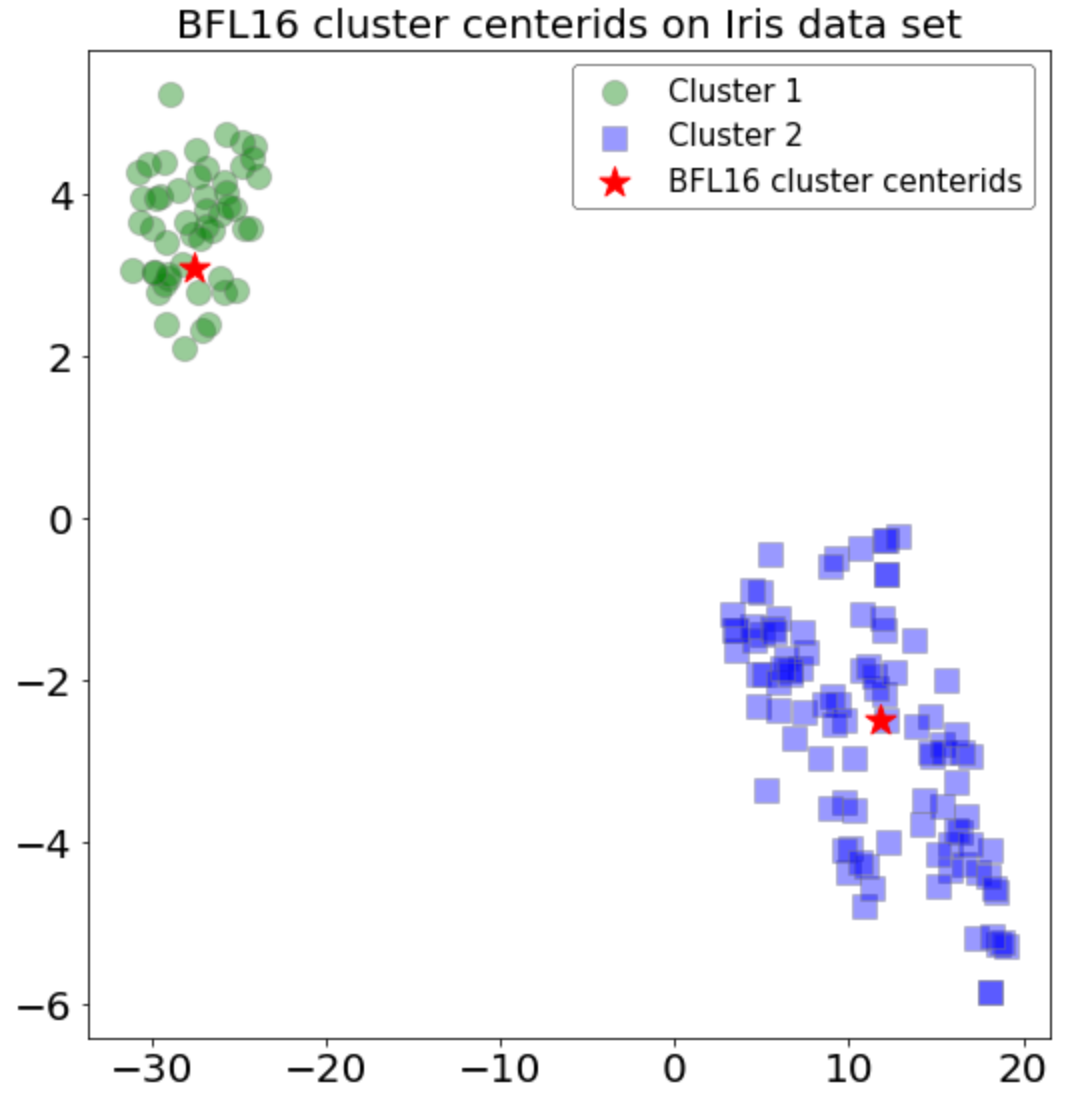}}
 \vspace{3pt}
 \end{minipage}
 \begin{minipage}{0.32\linewidth}
 \vspace{3pt}
 \centerline{\includegraphics[width=\textwidth]{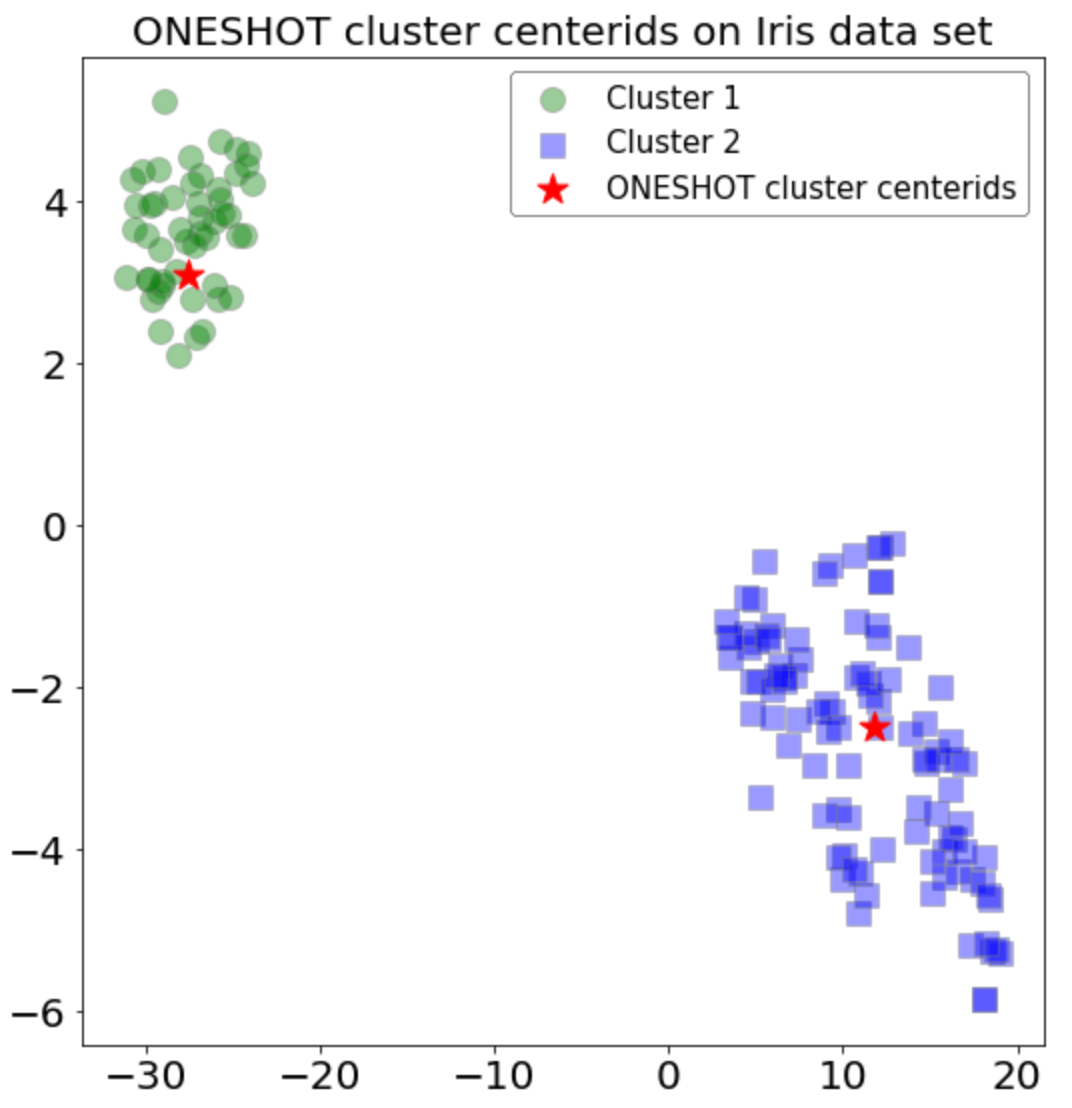}}
 \vspace{3pt}
 \end{minipage}
 \caption{A comparison of two quantum clusters and classical cluster centroids on Iris data set}
 \label{qvc-Iris}
\end{figure}

\begin{figure}[h]
 \begin{minipage}{0.32\linewidth}
 \vspace{3pt}
 \centerline{\includegraphics[width=\textwidth]{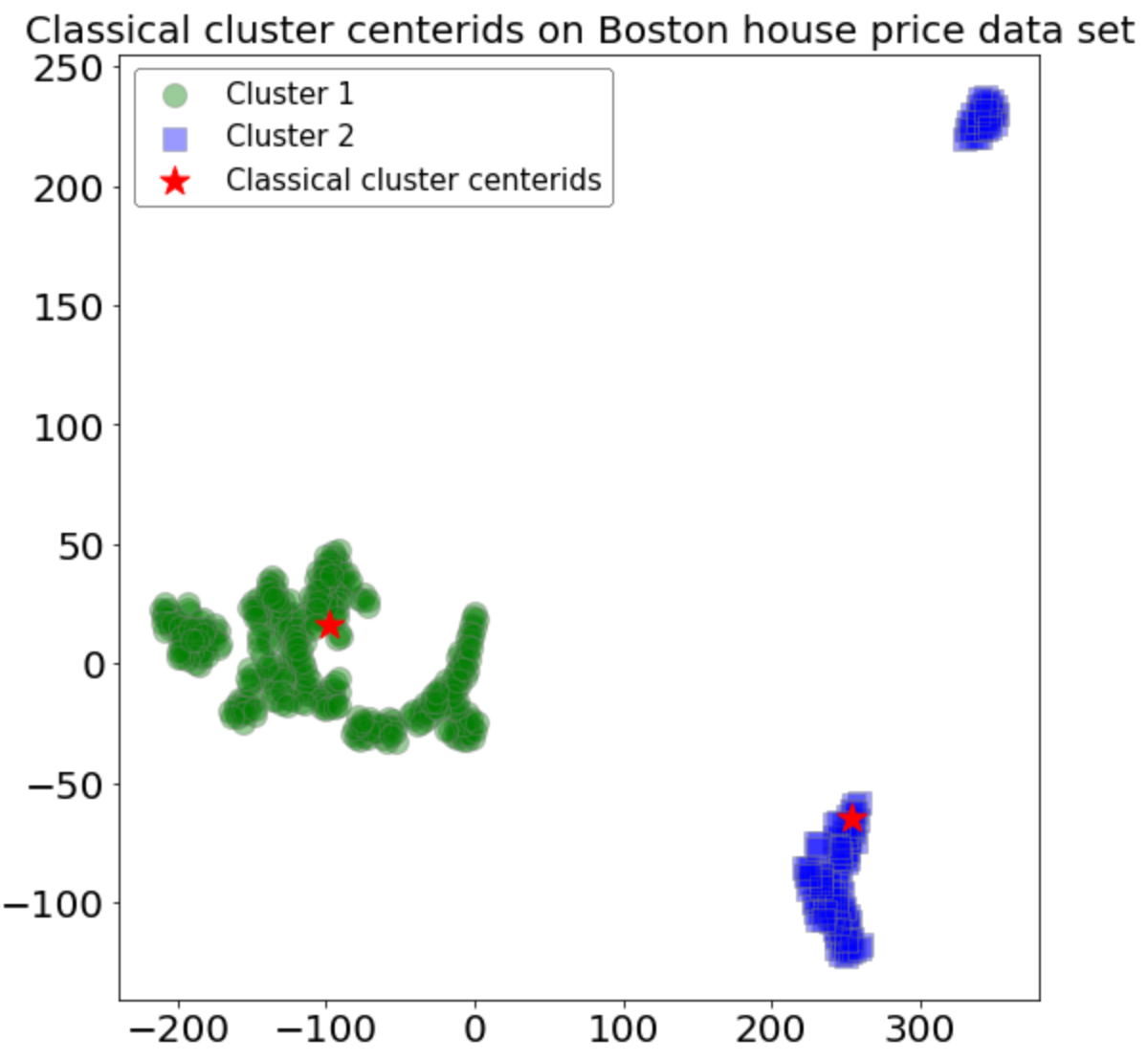}}
 \vspace{3pt}
 \end{minipage}
 \begin{minipage}{0.32\linewidth}
 \vspace{3pt}
 \centerline{\includegraphics[width=\textwidth]{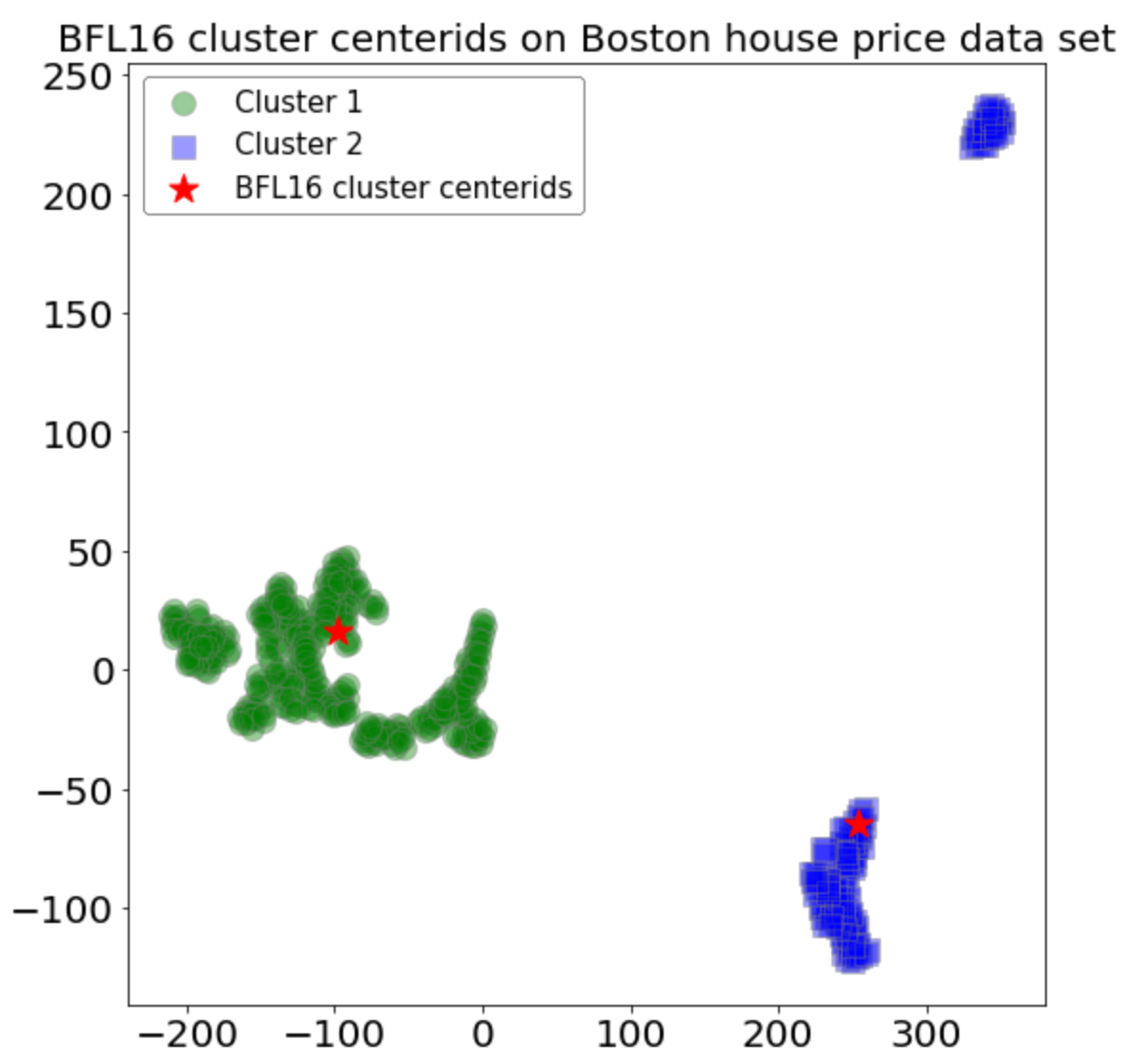}}
 \vspace{3pt}
 \end{minipage}
 \begin{minipage}{0.32\linewidth}
 \vspace{3pt}
 \centerline{\includegraphics[width=\textwidth]{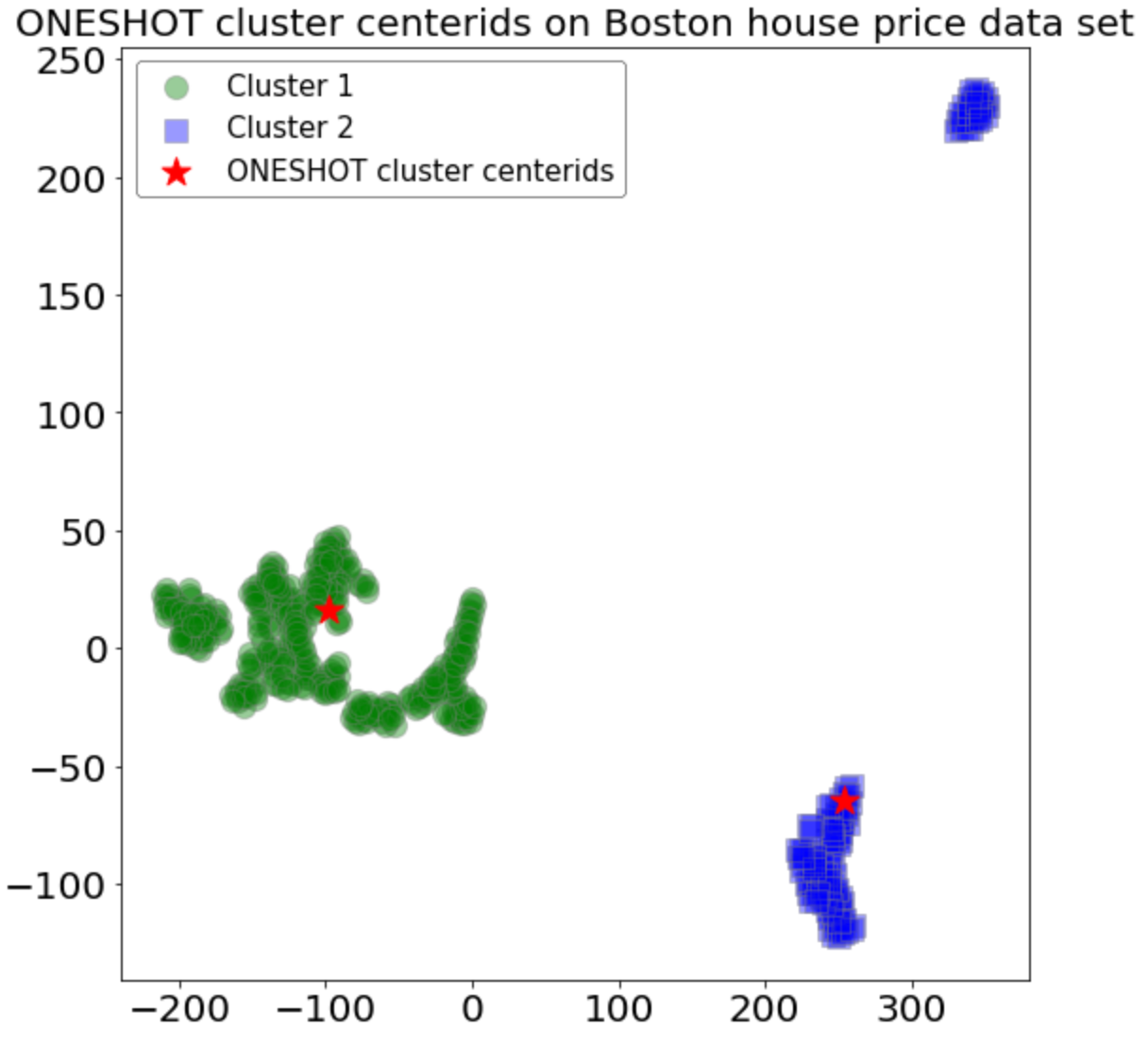}}
 \vspace{3pt}
 \end{minipage}
 \caption{A comparison of two quantum clusters and classical cluster centroids on Boston data set}
 \label{qvc-Boston}
 \end{figure}

\begin{figure}[h]
 \begin{minipage}{0.32\linewidth}
 \vspace{3pt}
 \centerline{\includegraphics[width=\textwidth]{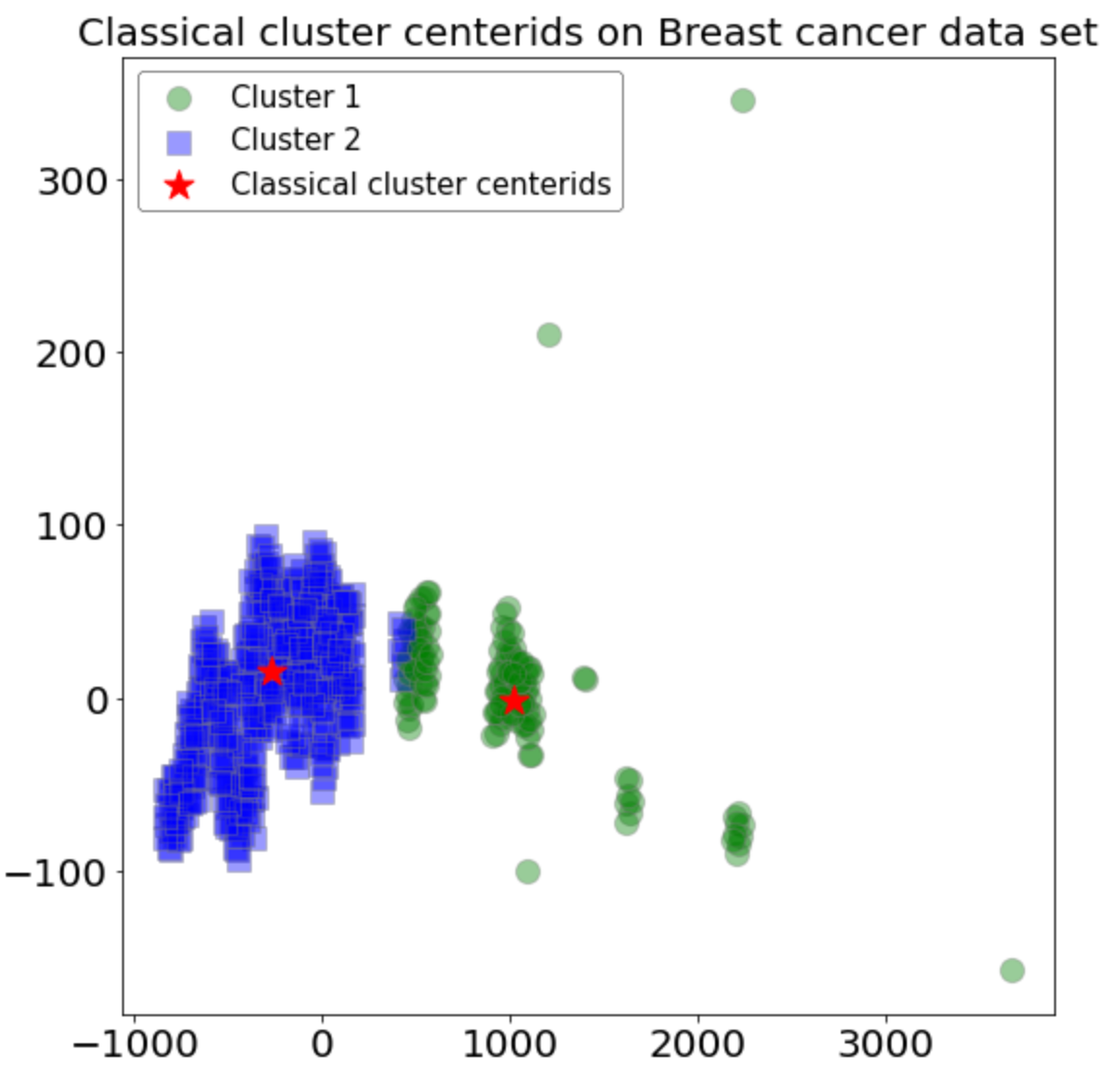}}
 \vspace{3pt}
 \end{minipage}
 \begin{minipage}{0.32\linewidth}
 \vspace{3pt}
 \centerline{\includegraphics[width=\textwidth]{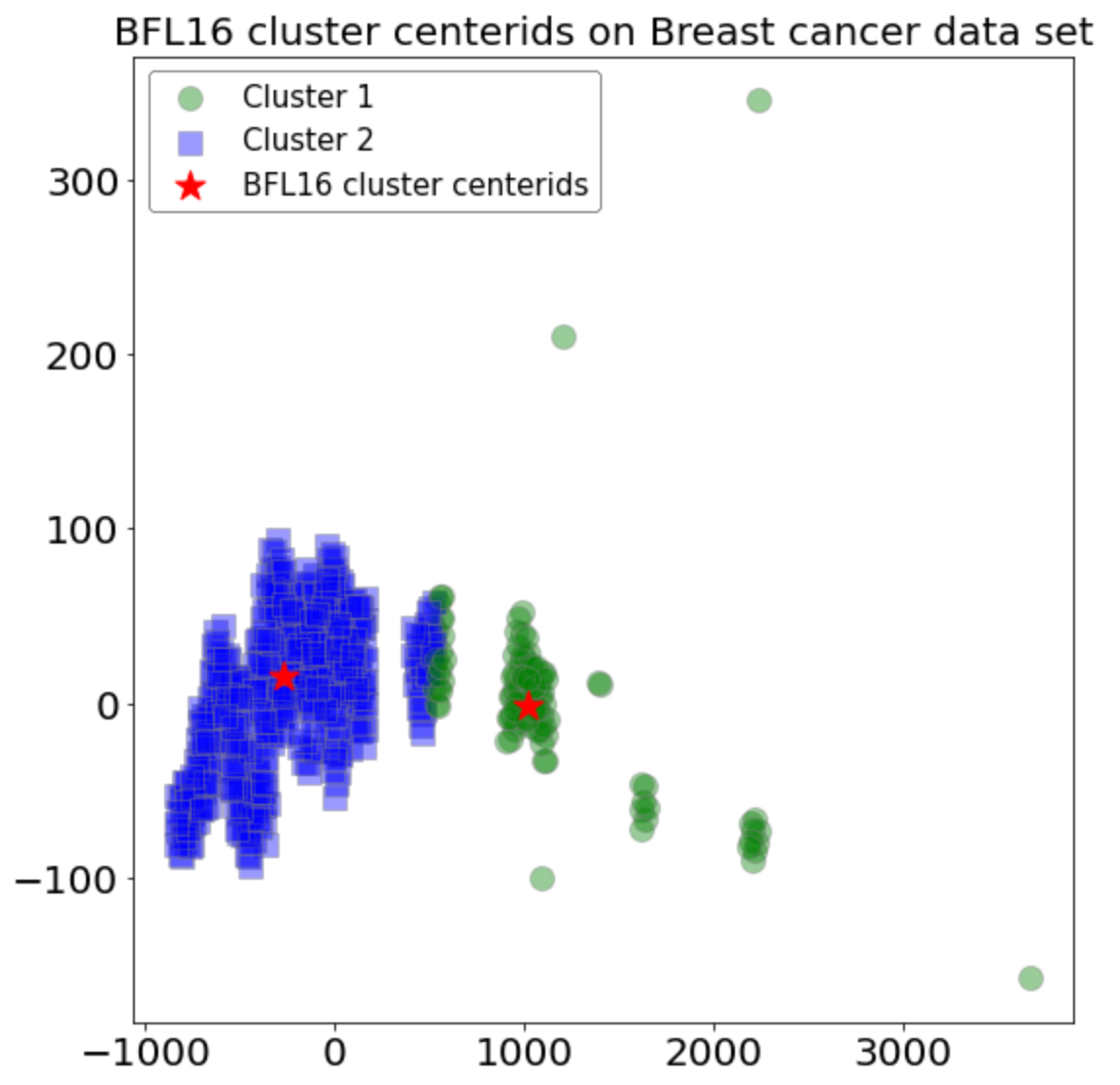}}
 \vspace{3pt}
 \end{minipage}
 \begin{minipage}{0.32\linewidth}
 \vspace{3pt}
 \centerline{\includegraphics[width=\textwidth]{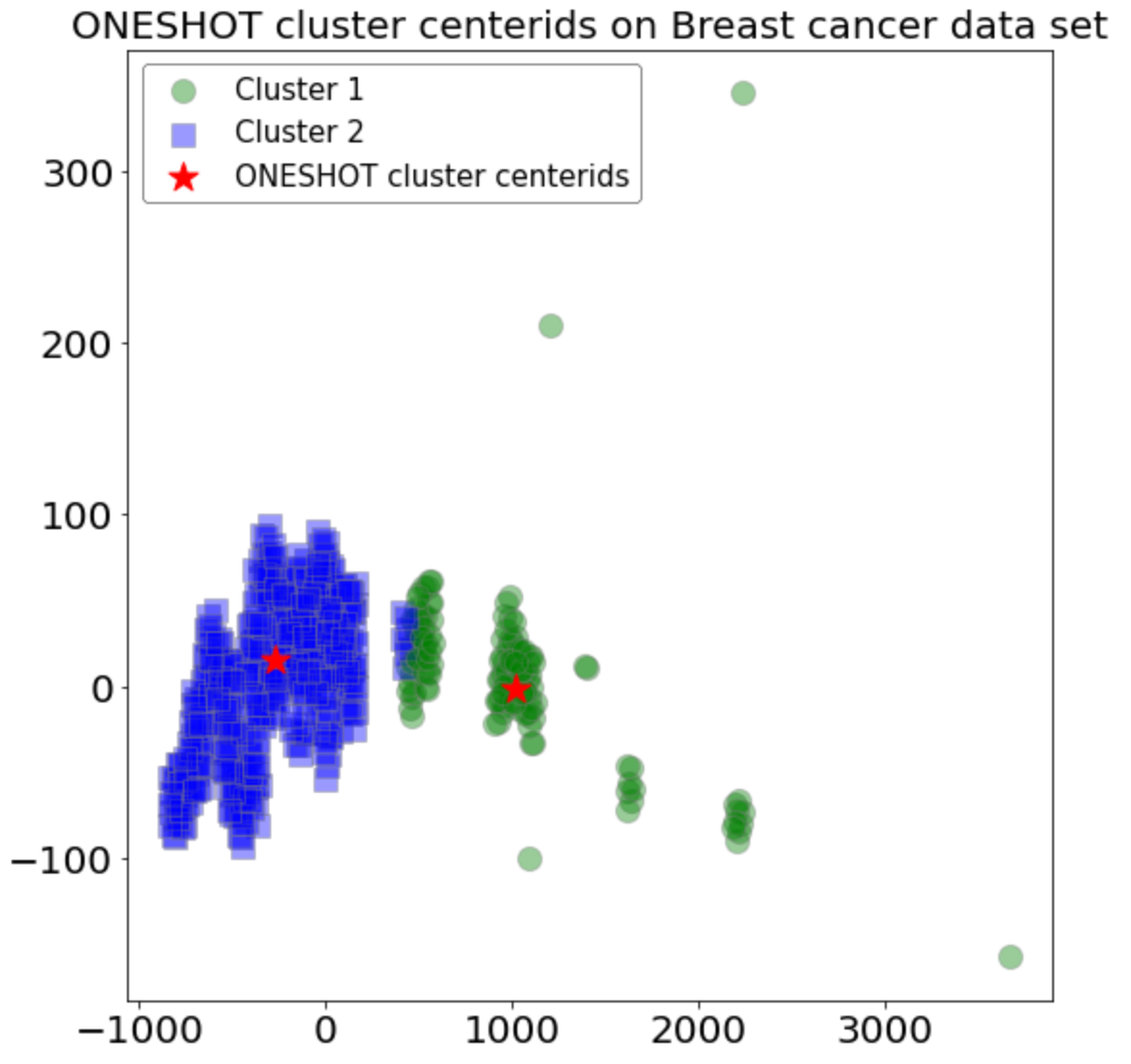}}
 \vspace{3pt}
 \end{minipage}
 \caption{A comparison of two quantum clusters and classical cluster centroids on Breast cancer data set}
 \label{qvc-Breast}
 \end{figure}

\begin{figure}[h]
 \begin{minipage}{0.32\linewidth}
 \vspace{3pt}
 \centerline{\includegraphics[width=\textwidth]{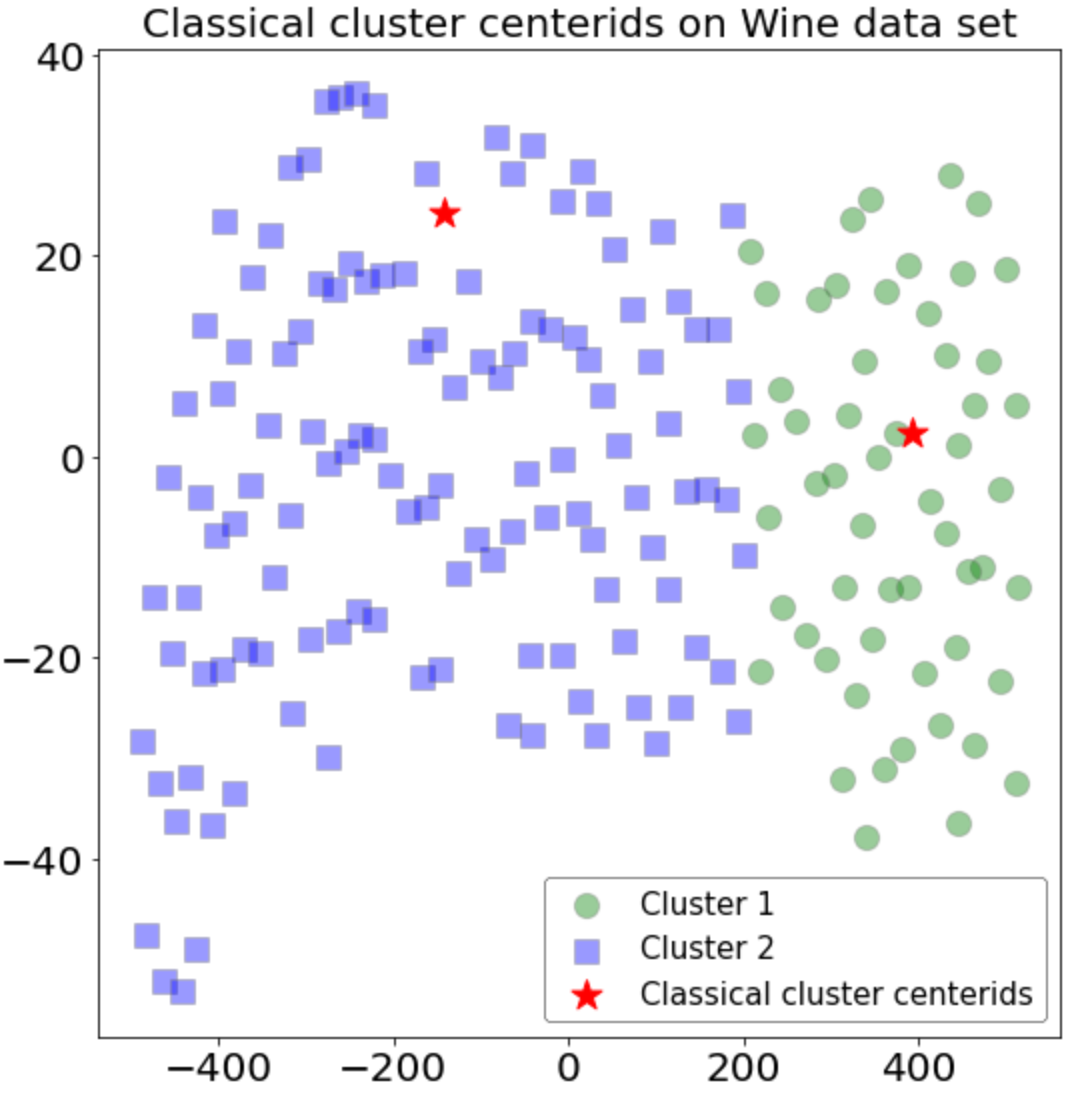}}
 \vspace{3pt}
 \end{minipage}
 \begin{minipage}{0.32\linewidth}
 \vspace{3pt}
 \centerline{\includegraphics[width=\textwidth]{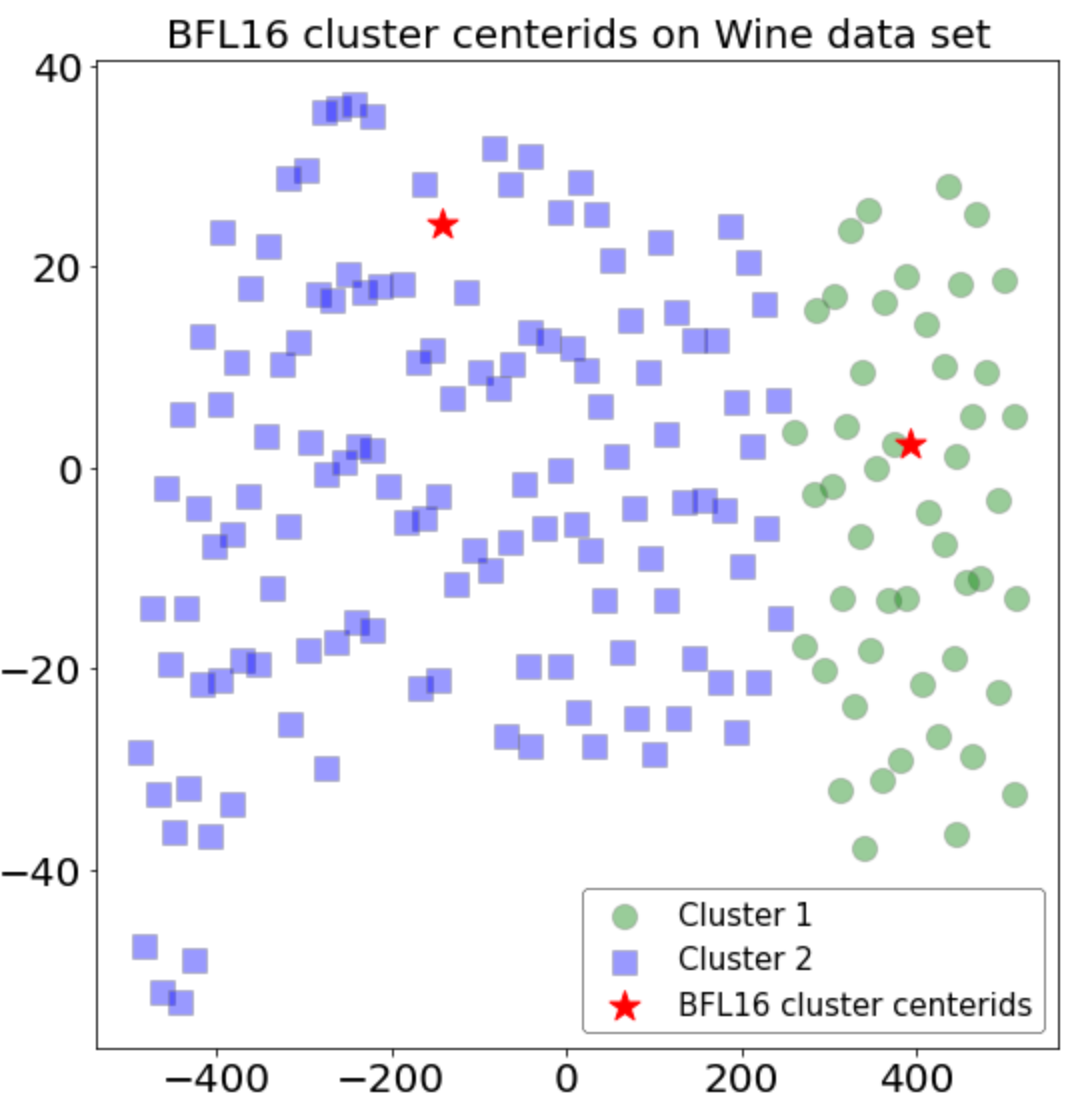}}
 \vspace{3pt}
 \end{minipage}
 \begin{minipage}{0.32\linewidth}
 \vspace{3pt}
 \centerline{\includegraphics[width=\textwidth]{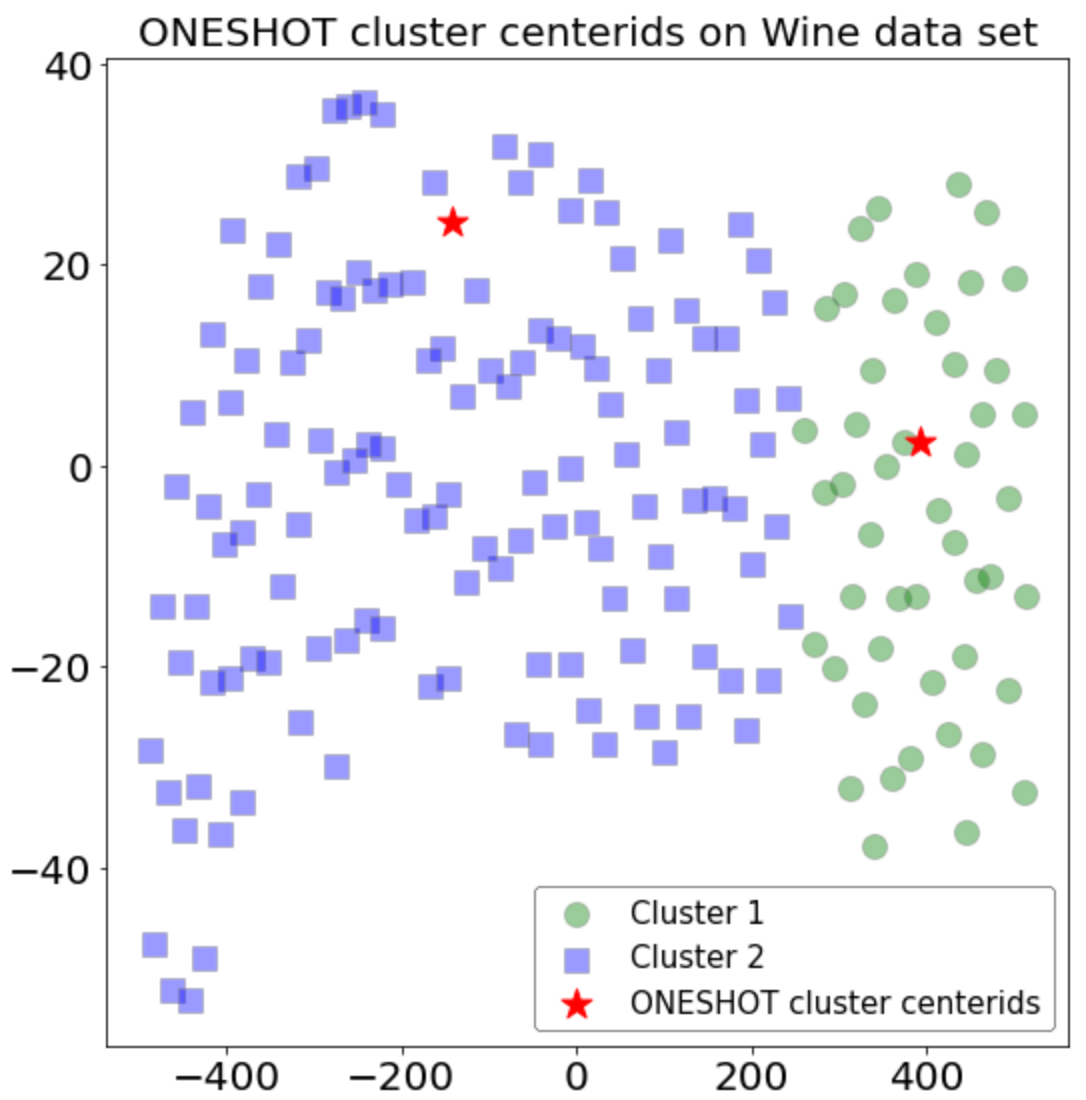}}
 \vspace{3pt}
 \end{minipage}
 \caption{A comparison of two quantum clusters and classical cluster centroids on Wine data set}
 \label{qvc-Wine}
 \end{figure}

\begin{figure}[h]
 \begin{minipage}{0.32\linewidth}
 \vspace{3pt}
 \centerline{\includegraphics[width=\textwidth]{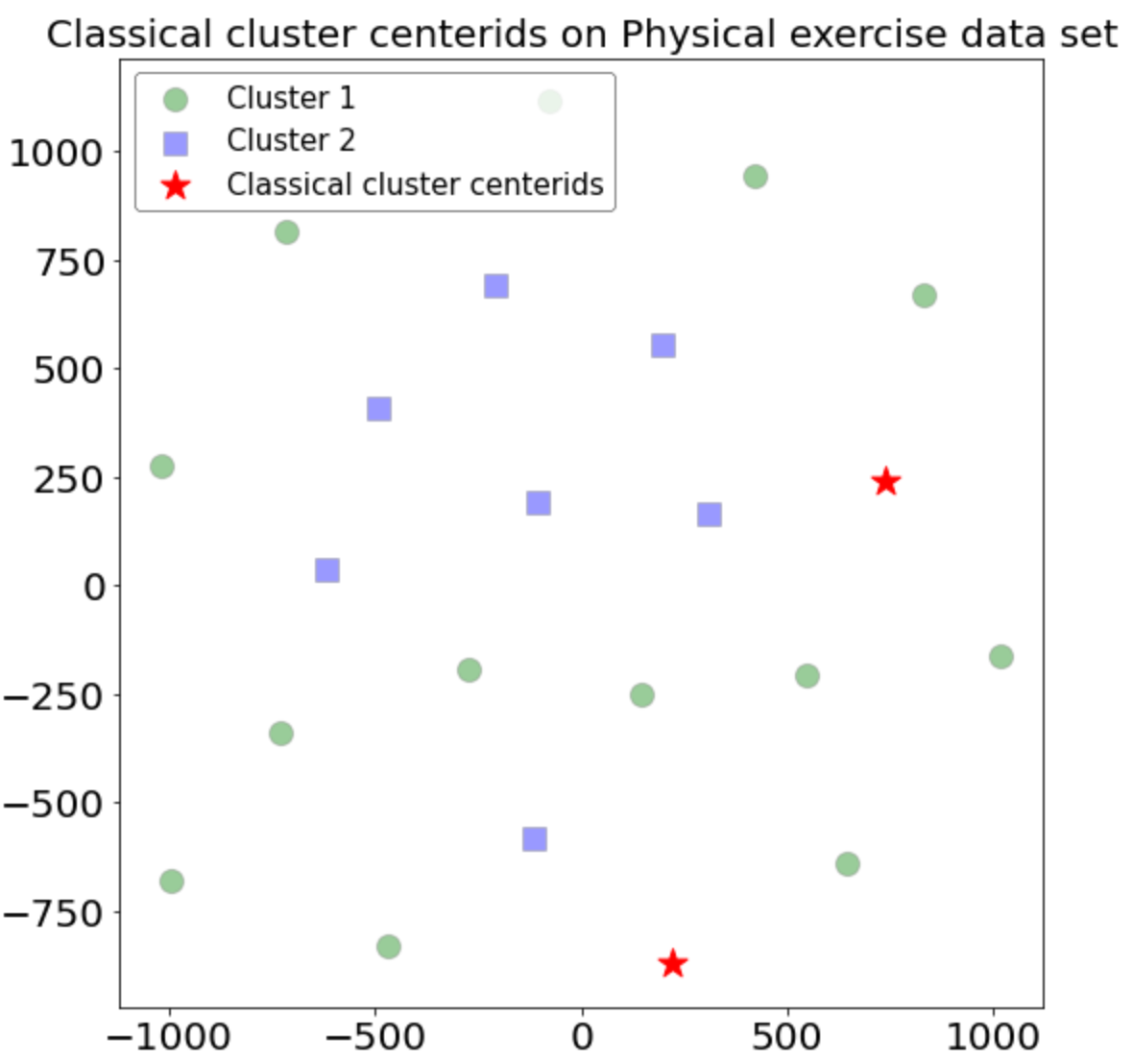}}
 \vspace{3pt}
 \end{minipage}
 \begin{minipage}{0.32\linewidth}
 \vspace{3pt}
 \centerline{\includegraphics[width=\textwidth]{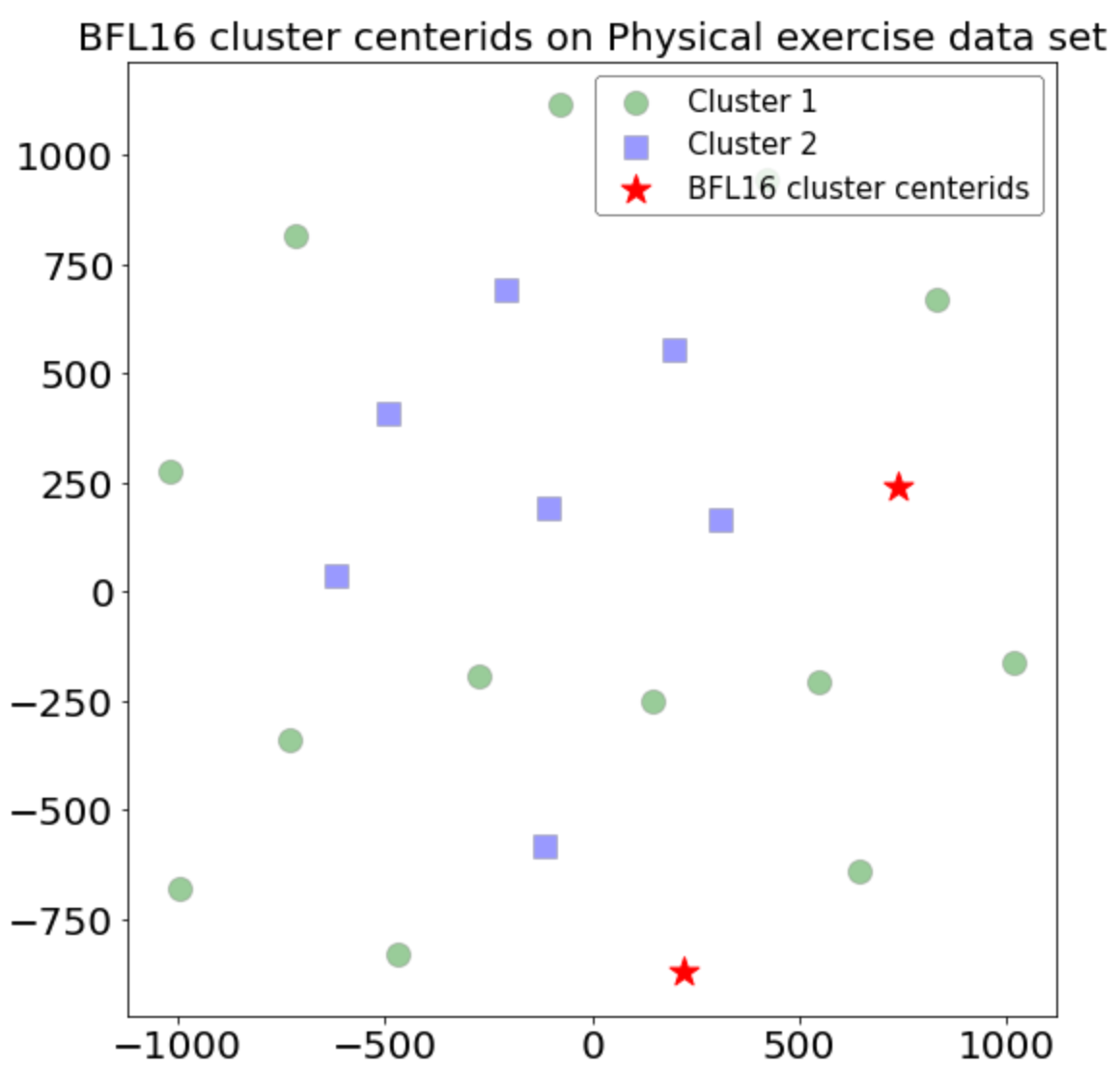}}
 \vspace{3pt}
 \end{minipage}
 \begin{minipage}{0.32\linewidth}
 \vspace{3pt}
 \centerline{\includegraphics[width=\textwidth]{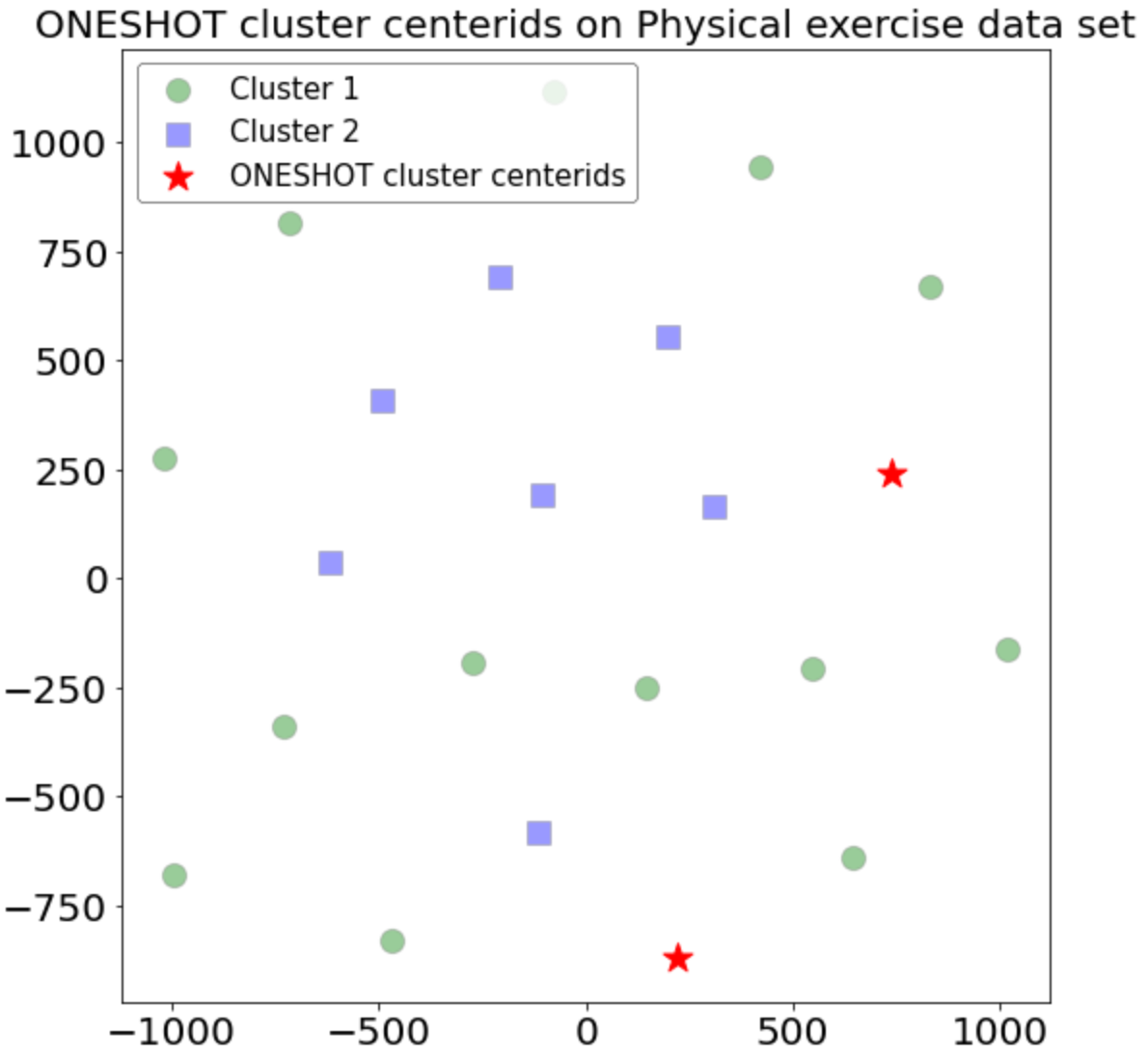}}
 \vspace{3pt}
 \end{minipage}
 \caption{A comparison of two quantum clusters and classical cluster centroids on Physical data set}
 \label{qvc-Physical}
 \end{figure}

\begin{figure}[h]
 \begin{minipage}{0.32\linewidth}
 \vspace{3pt}
 \centerline{\includegraphics[width=\textwidth]{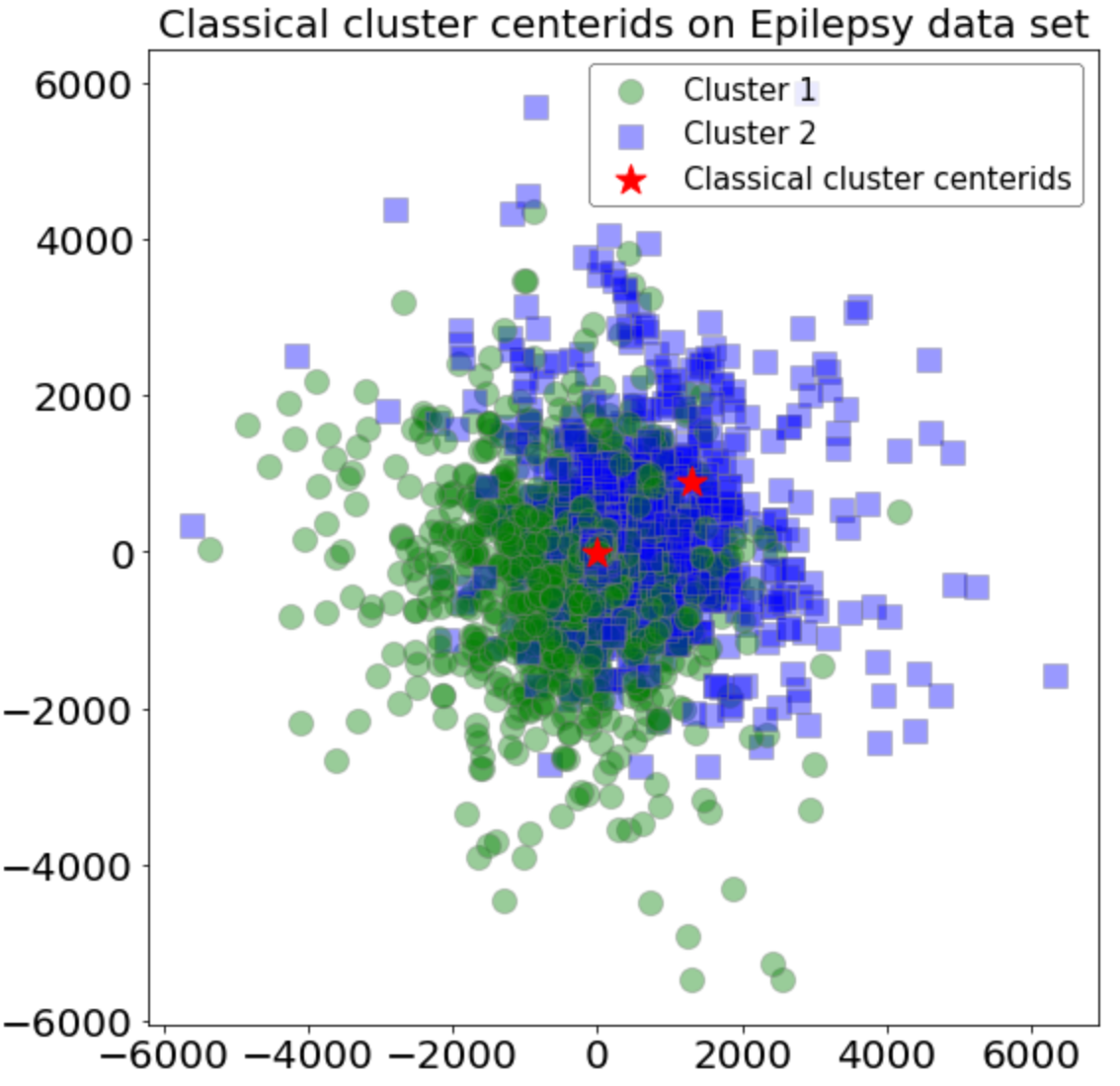}}
 \vspace{3pt}
 \end{minipage}
 \begin{minipage}{0.32\linewidth}
 \vspace{3pt}
 \centerline{\includegraphics[width=\textwidth]{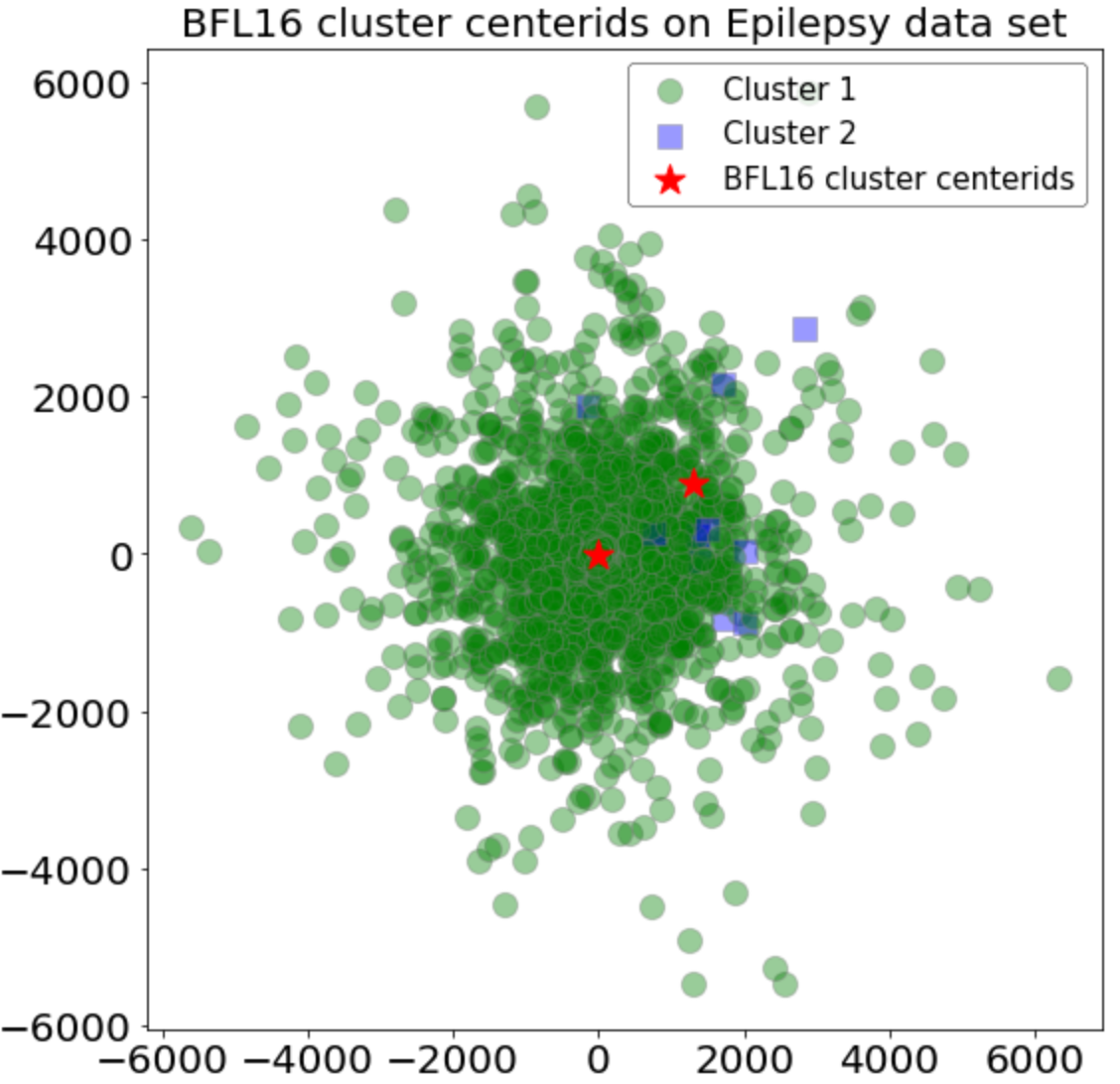}}
 \vspace{3pt}
 \end{minipage}
 \begin{minipage}{0.32\linewidth}
 \vspace{3pt}
 \centerline{\includegraphics[width=\textwidth]{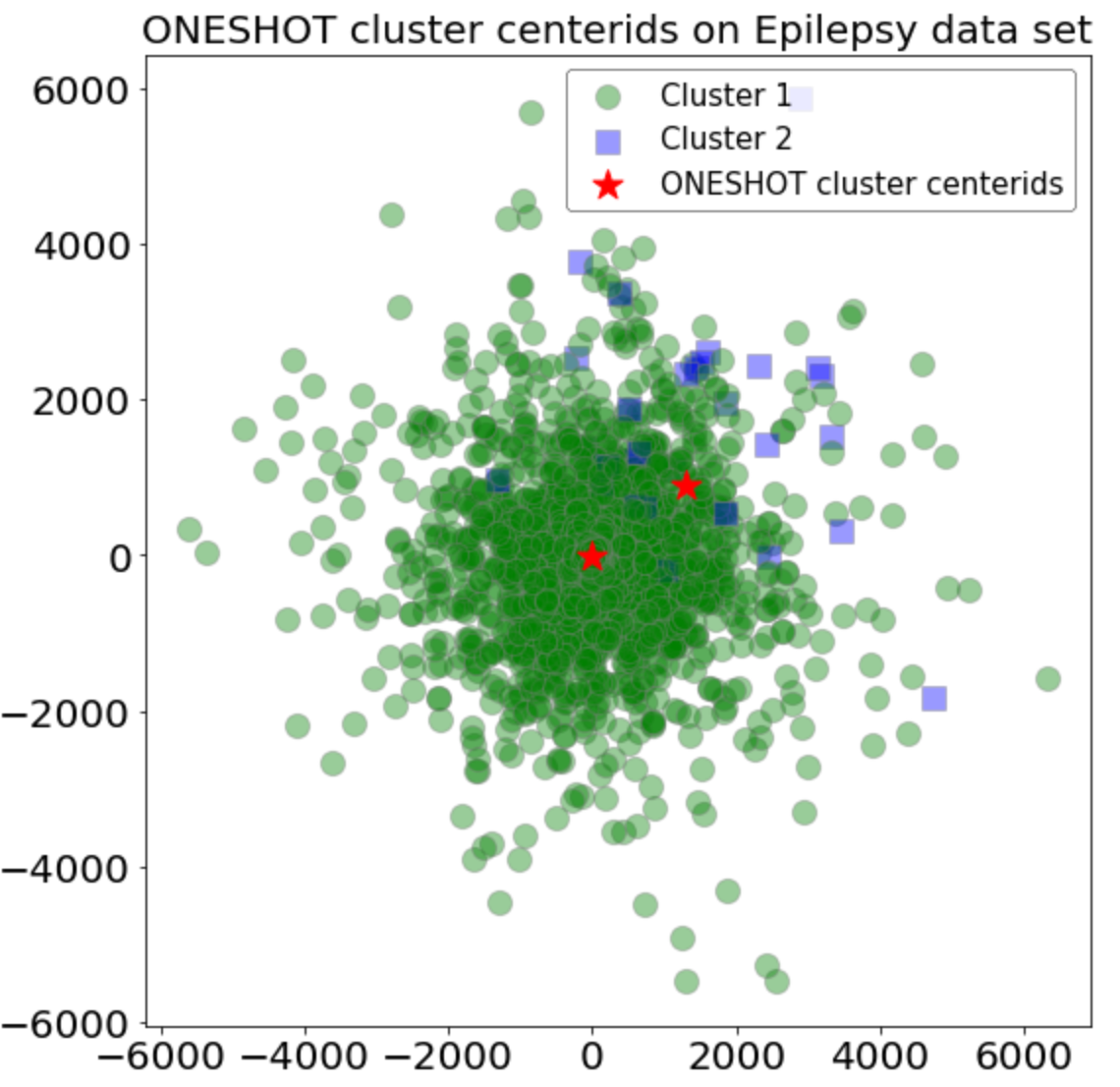}}
 \vspace{3pt}
 \end{minipage}
 \caption{A comparison of two quantum clusters and classical cluster centroids on Epilepsy data set}
 \label{qvc-Epilepsy}
 \end{figure}

Referring to this approach, the quantum cluster centroids for 6 data sets are plotted in Figure \ref{qvc-Iris}, Figure \ref{qvc-Boston}, Figure \ref{qvc-Breast}, Figure \ref{qvc-Wine}, Figure \ref{qvc-Physical} and Figure \ref{qvc-Epilepsy} respectively. The two clusters are in green dots and blue squares separately, and the quantum cluster centroids are shown in the two figures on the right, and the classical cluster centroids on the left. For all 6 data sets, quantum cluster centroids and classical cluster centroids are very close, which means that quantum K-Means clustering can make clusters that are almost the same as classical K-Means clustering. 


\subsection{Coreset size} \label{csize}

\begin{figure}[h]
 \begin{minipage}{0.48\linewidth}
 \vspace{3pt}
 \centerline{\includegraphics[width=\textwidth]{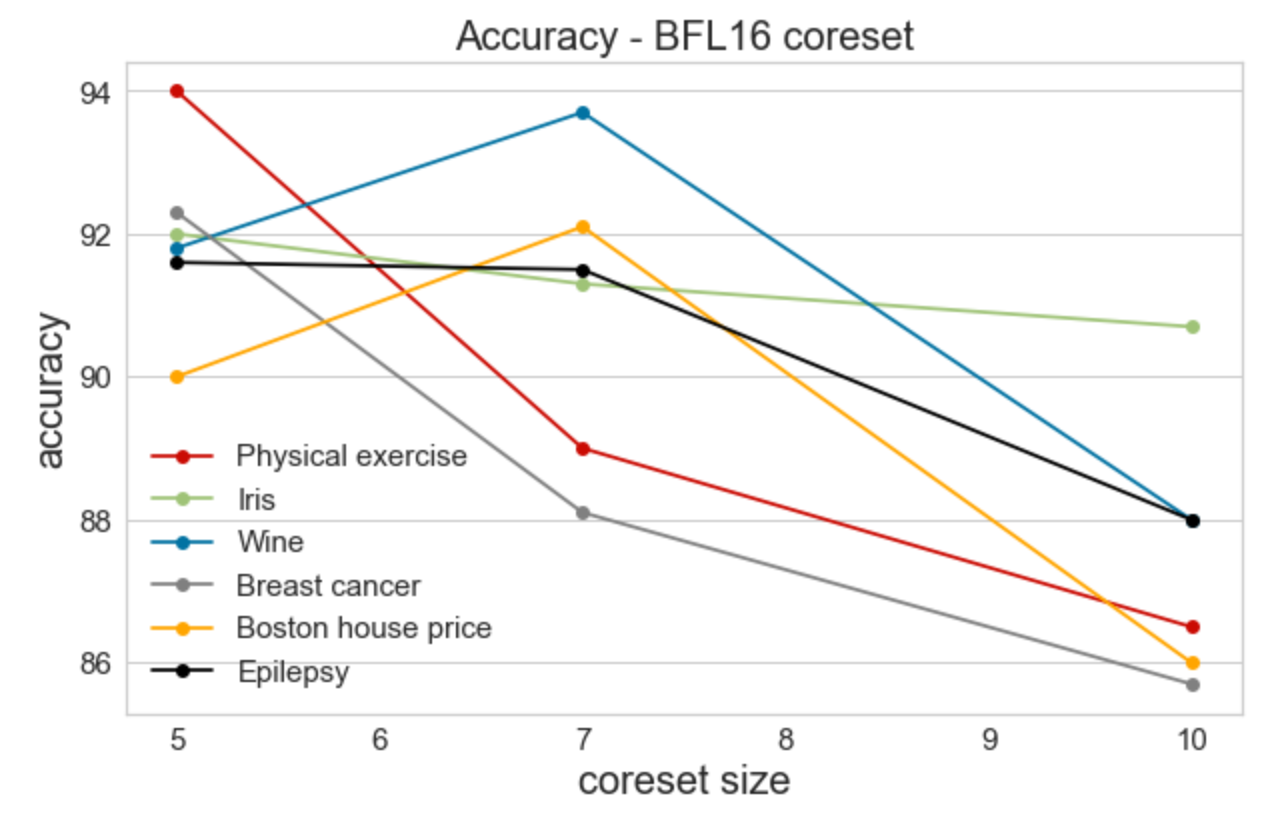}}
 \centerline{\includegraphics[width=\textwidth]{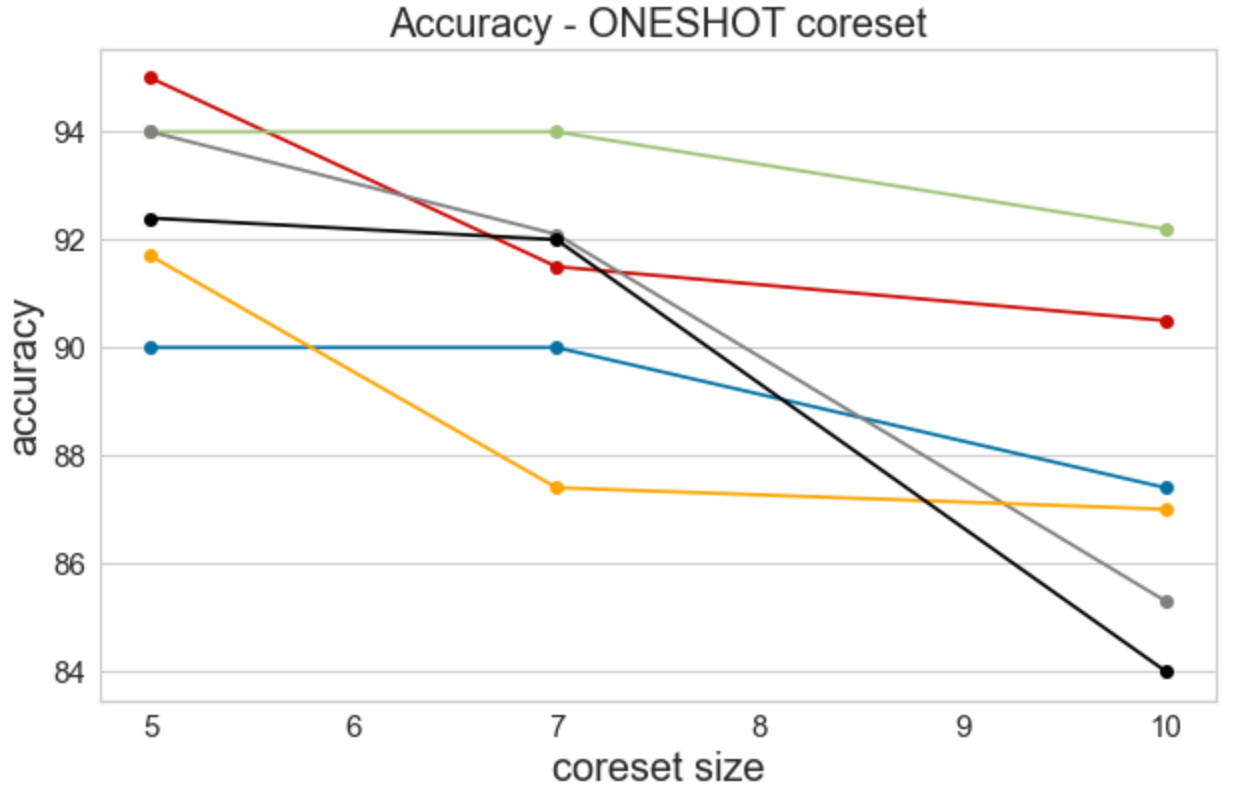}}
 \vspace{3pt}
 \end{minipage}
 \begin{minipage}{0.48\linewidth}
 \vspace{3pt}
 \centerline{\includegraphics[width=\textwidth]{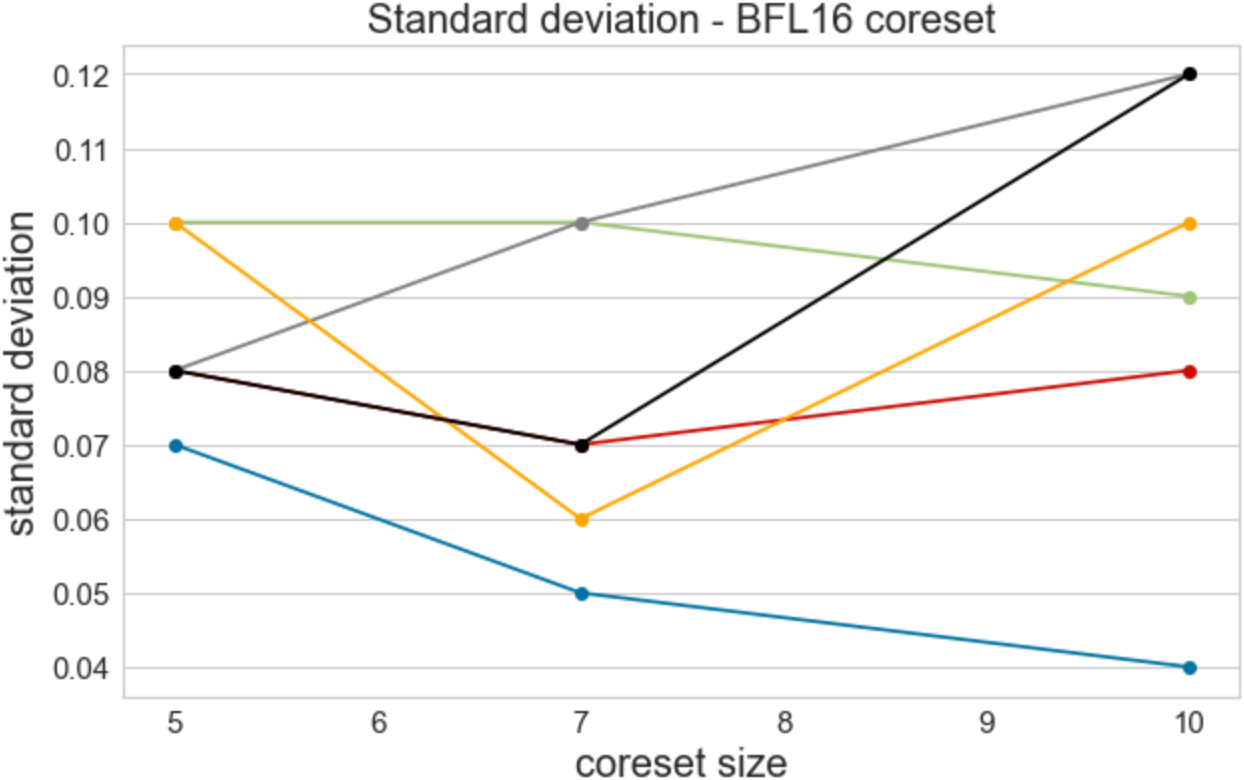}}
 \centerline{\includegraphics[width=\textwidth]{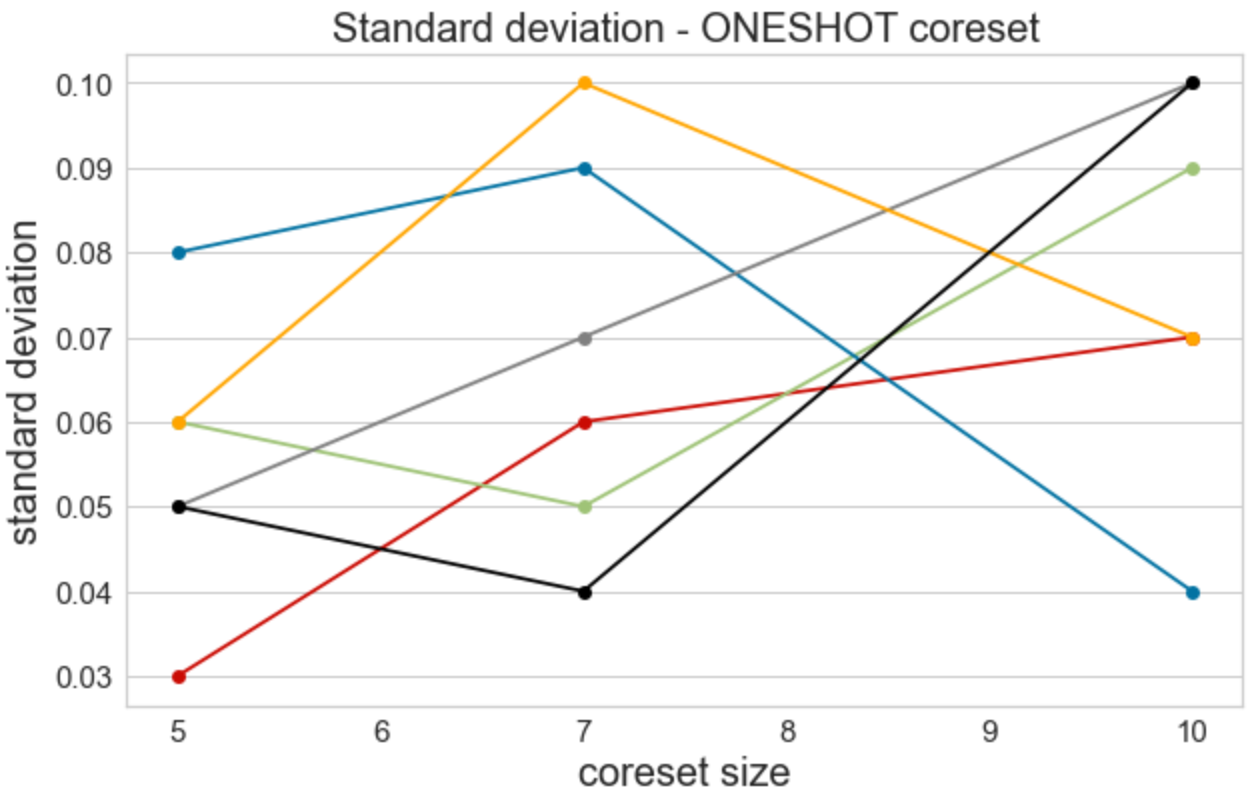}}
 \vspace{3pt}
 \end{minipage}
 \caption{Relationship between coreset size and accuracy}
 \label{quantum coreset size-accuracy}
 \end{figure}

\begin{figure}[h]
 \begin{minipage}{0.32\linewidth}
 \vspace{3pt}
 \centerline{\includegraphics[width=\textwidth]{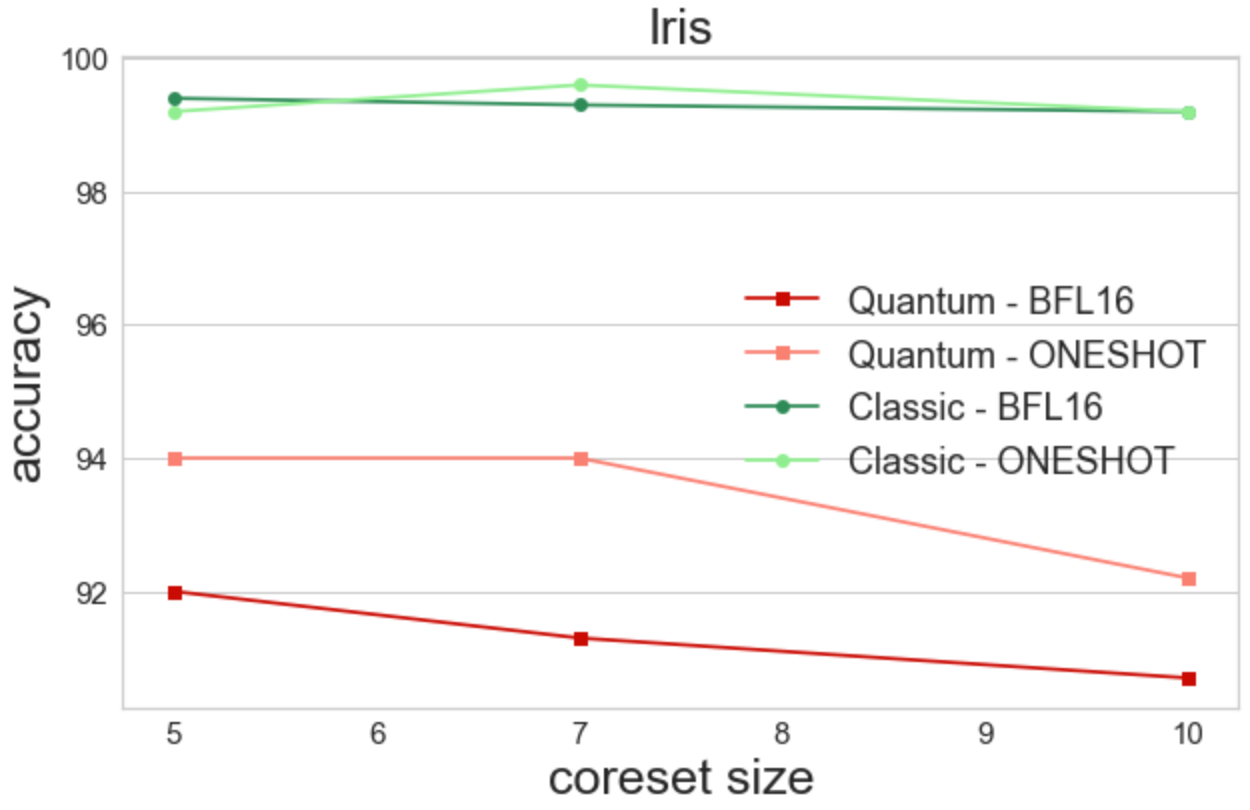}}
 \vspace{3pt}
 \centerline{\includegraphics[width=\textwidth]{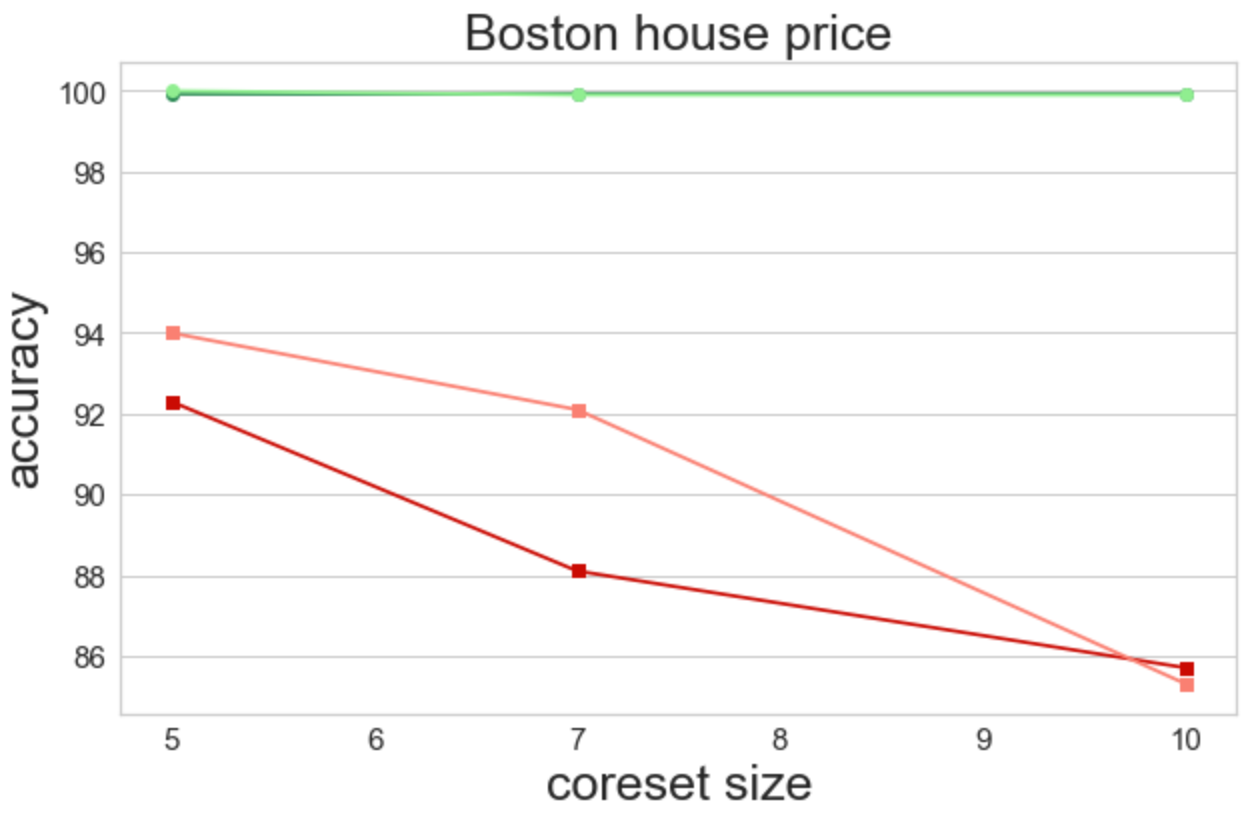}}
 \vspace{3pt}
 \end{minipage}
 \begin{minipage}{0.32\linewidth}
 \vspace{3pt}
 \centerline{\includegraphics[width=\textwidth]{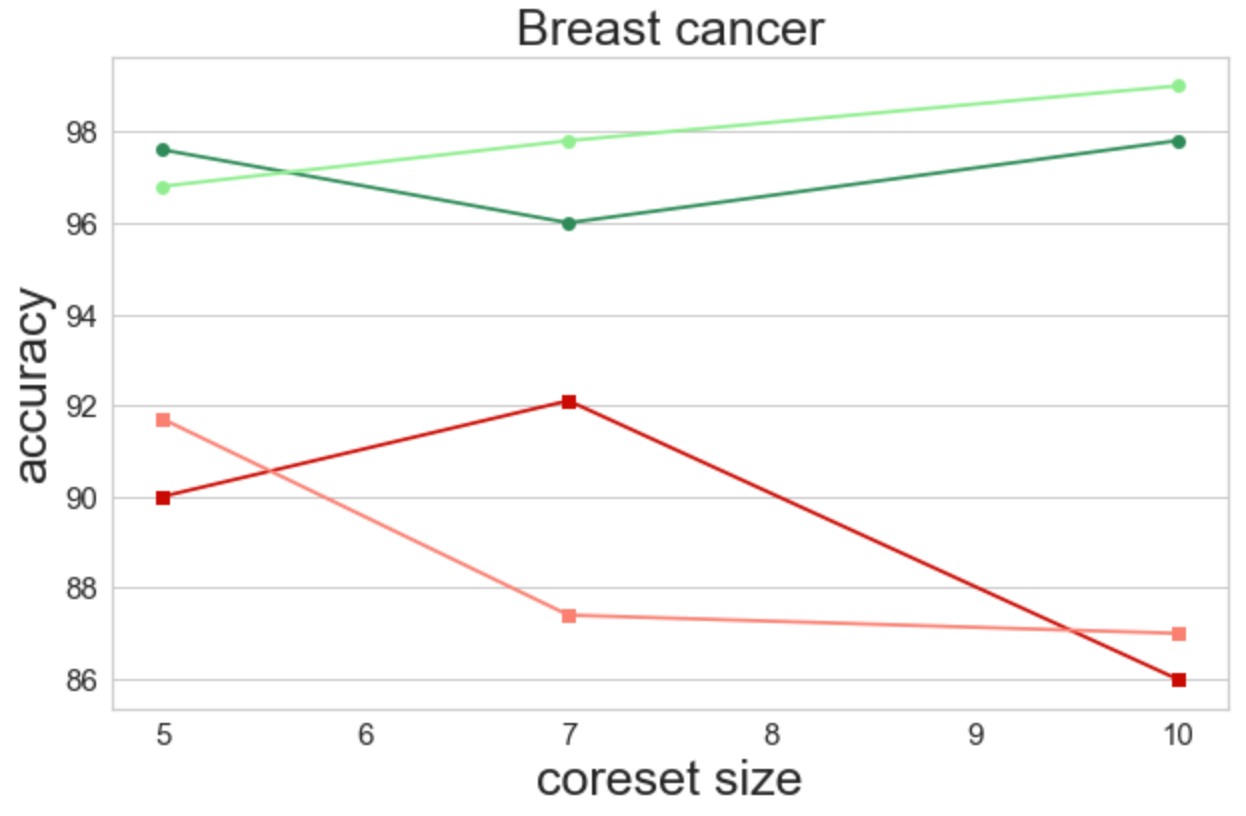}}
 \vspace{3pt}
 \centerline{\includegraphics[width=\textwidth]{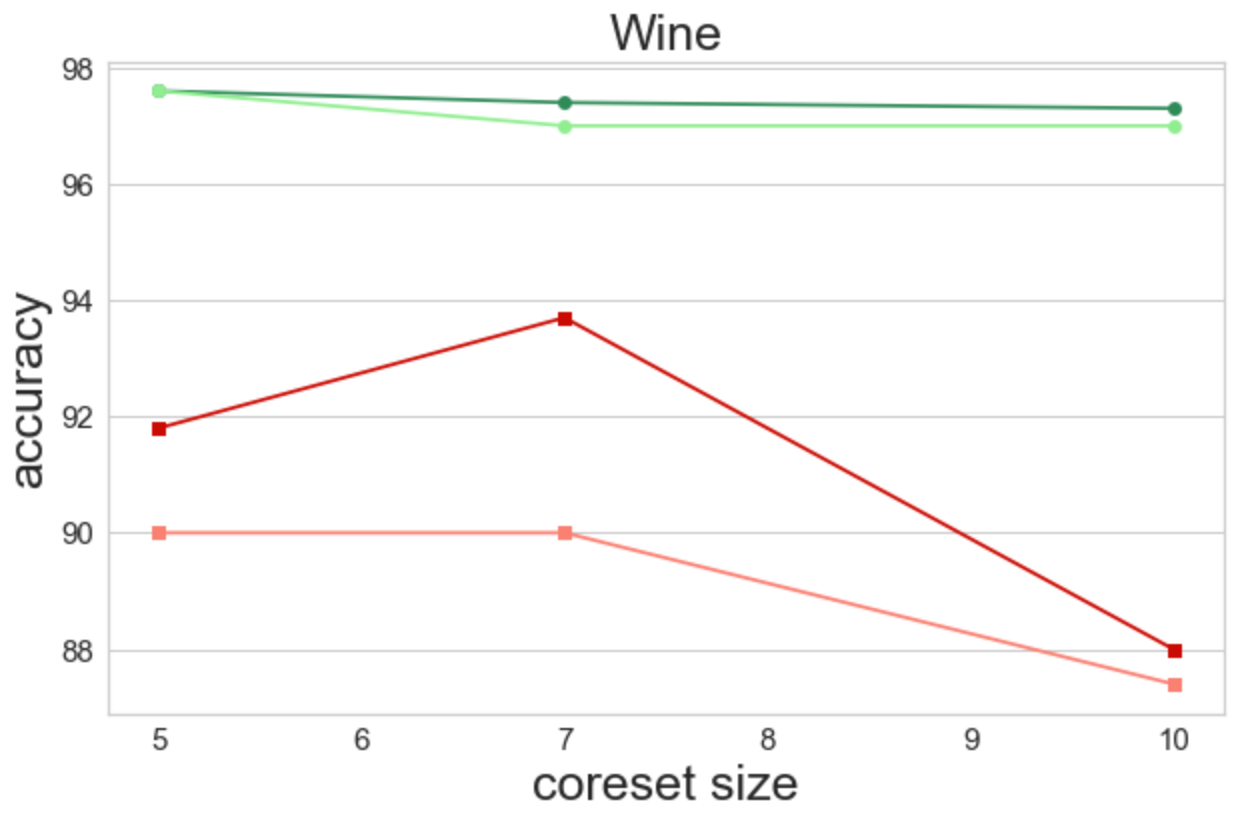}}
 \vspace{3pt}
 \end{minipage}
 \begin{minipage}{0.32\linewidth}
 \vspace{3pt}
 \centerline{\includegraphics[width=\textwidth]{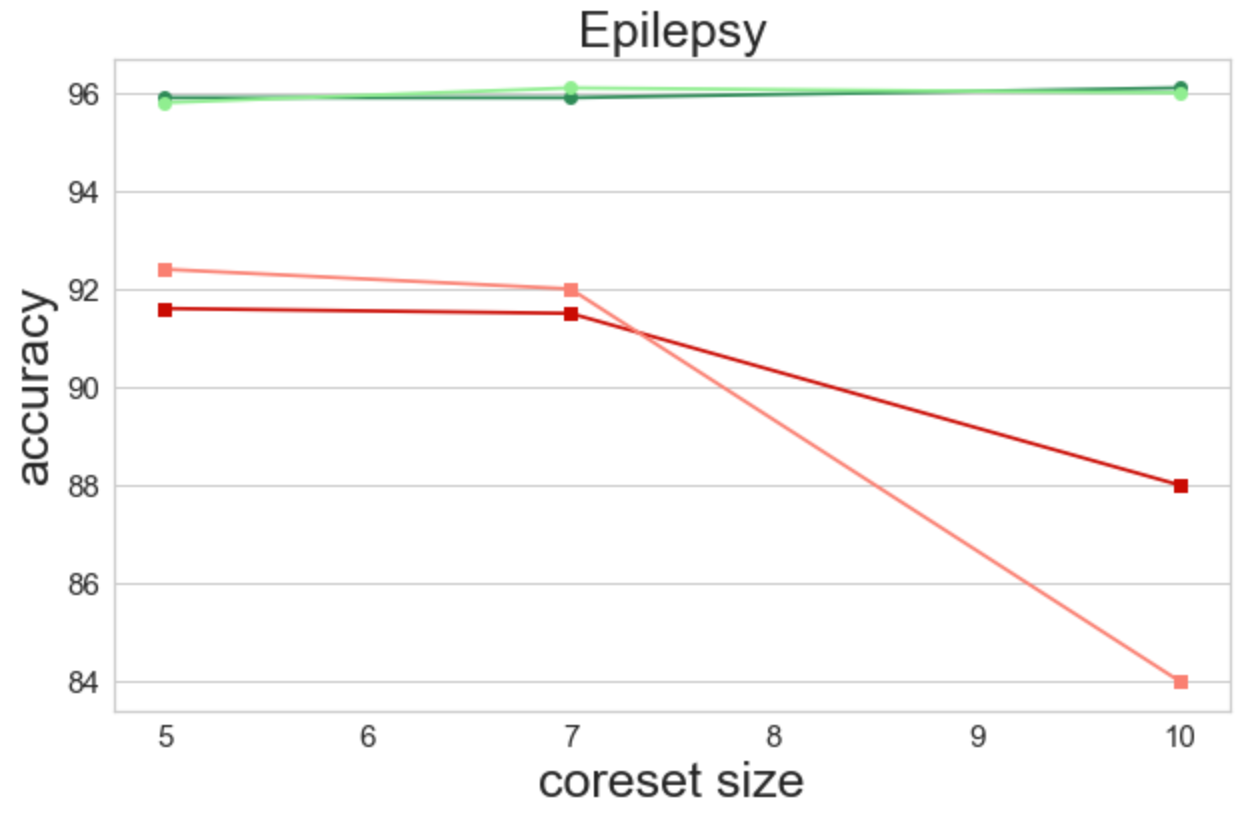}}
 \vspace{3pt}
 \centerline{\includegraphics[width=\textwidth]{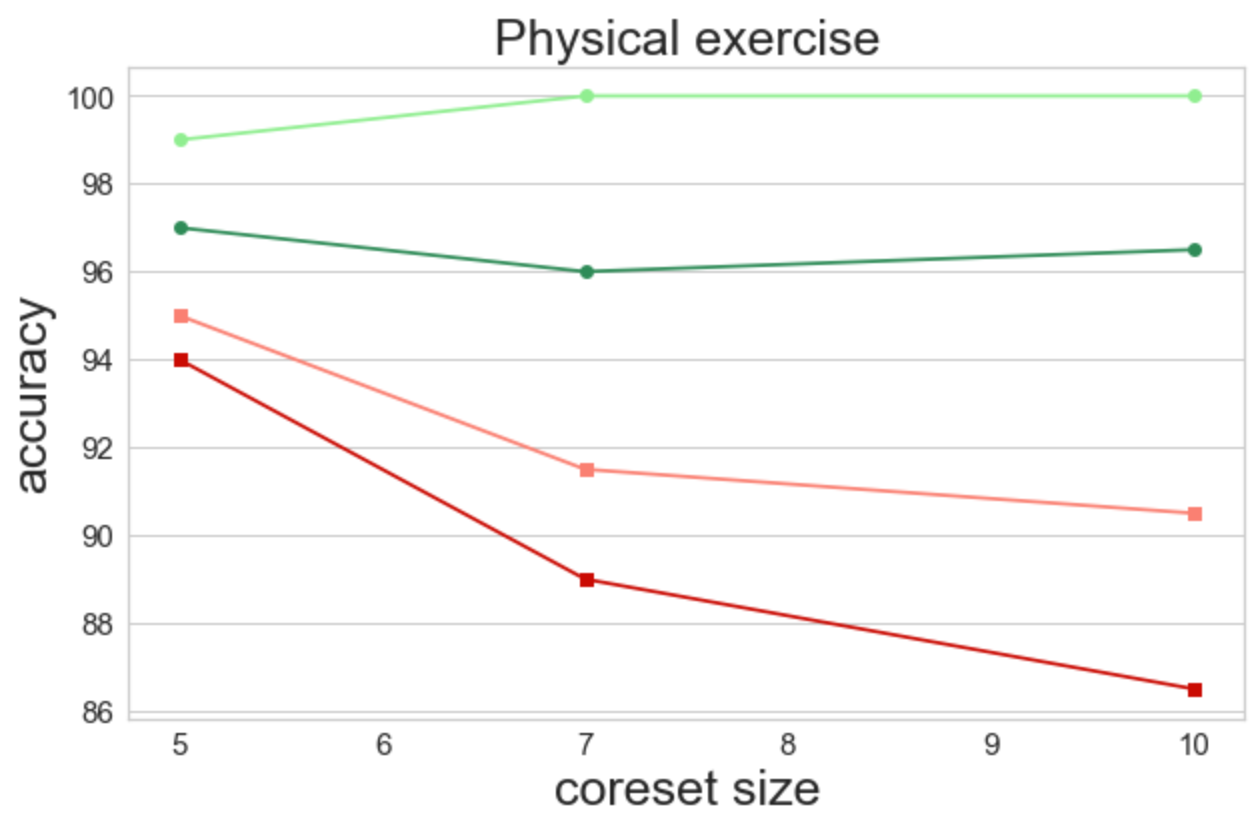}}
 \vspace{3pt}
 \end{minipage}
 \caption{classical accuracy - quantum accuracy}
 \label{c-q accuracy}
 \end{figure}

The coreset size has a significant influence on quantum K-Means clustering. Recent work \cite{tomesh2020coreset} showed a uniform trend for all six data sets, that is, the larger the size of the coreset, the less the accuracy loss. A key question is does a larger coreset size help quantum K-Means clustering more? The result of experiments shows that a larger coreset gives a negative impact on accuracy. The relationship between accuracy and coreset size using QAOA and classical segmentation and clustering operations are plotted to support, as shown in Figure \ref{c-q accuracy}. For the six data sets in Table \ref{Datasets}, whether it is the BFL16 coreset or the ONESHOT coreset when using the classical method, the accuracy increases with the increase of the coreset size, but when using QAOA with Nelder-Mead optimizer, the accuracy decreases with the increase of the coreset size, $i.e$. coreset size will directly affect the performance of QAOA with the Nelder-Mead optimizer, which in turn has an impact on accuracy. Figure \ref{quantum coreset size-accuracy} illustrates the impact of coreset size on the standard deviation and accuracy of the six data sets. For both the BFL16 coreset and ONESHOT coreset, the overall trend is that the accuracy decreases as the coreset size increases, while the standard deviation increases. It is difficult for QAOA to find suitable quantum states to maximize the Hamiltonian if more quantum states are introduced. If the coreset size is 5, there will be 5 qubits used and  $2^5$ quantum states. The amount of quantum state increases exponentially with the number of qubits. The impact of coreset size on accuracy is also data-dependent. The accuracy increases slightly when the coreset size increases from 5 to 7 for the Boston house price data set and wine data set, but it drops sharply when the coreset size increases to 10, while other data sets give different accuracy variations. Accuracy on Epilepsy data set decreases more than other data sets when coreset size increases from 7 to 10, but remains stable when coreset size changes from 5 to 7.

An open research area is whether other optimizers can optimize the QAOA algorithm more efficiently or whether they also follow the above trend when the coreset size grows. These experiments are all done on the noise-free simulator. If on the real quantum hardware with quantum noise, the large circuit will suffer from more noise, so it is necessary to take this factor into account to make a compromise.

\subsection{Quantum noise and Quantum AutoEncoders}

If QAOA is disturbed by quantum noise, the final quantum result will be inaccurate, leading to wrong segmentation and clustering. The quantum cluster centroids from wrong segmentation and clustering will inaccurately summarize the data set and leads to low accuracy. We use a depolarizing noise model to introduce depolarisation quantum noise on all qubits of the simulator. 

\begin{figure}[h]
 \begin{minipage}{\linewidth}
 \vspace{3pt}
 \centerline{\includegraphics[width=\textwidth]{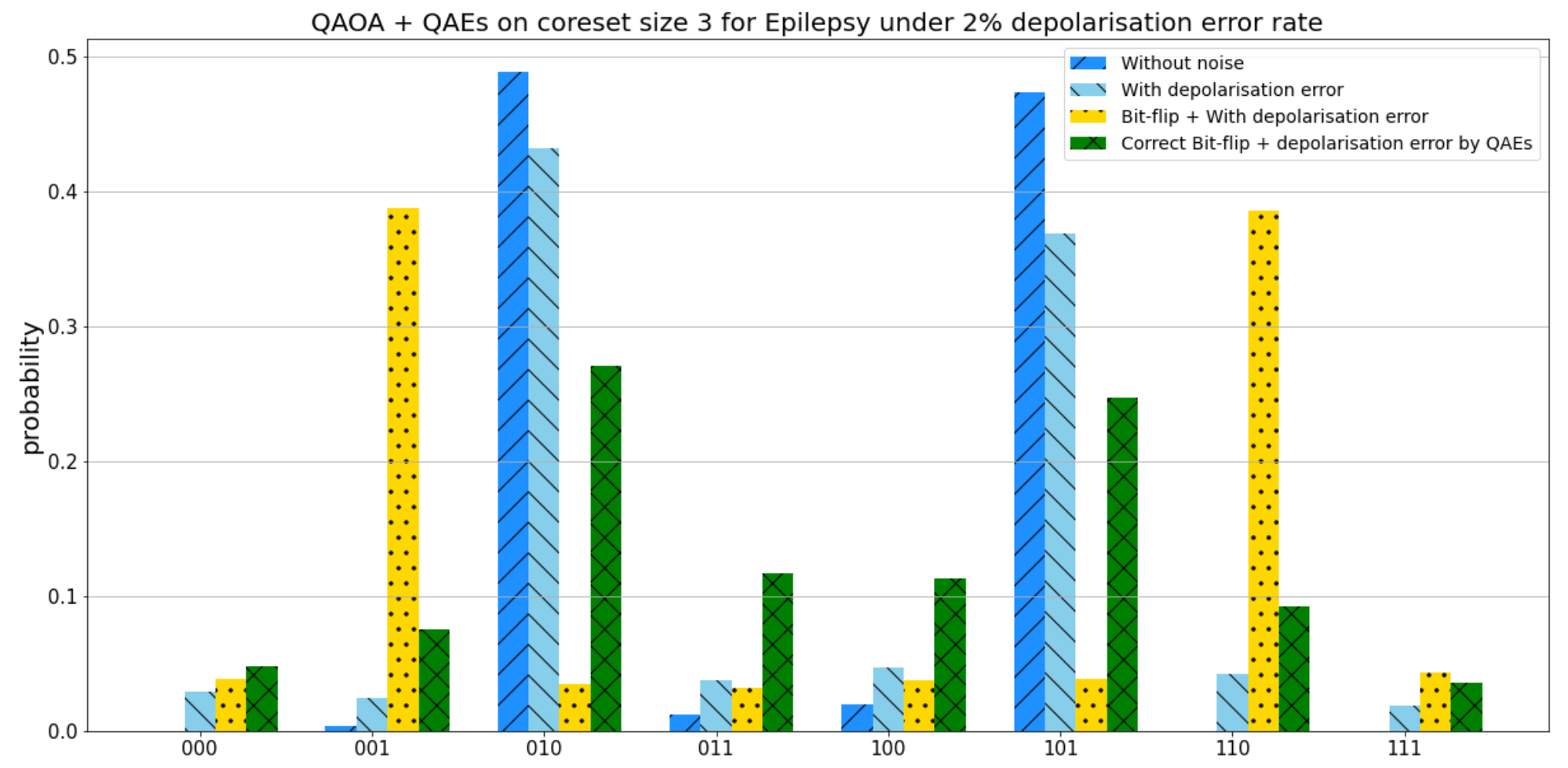}}
 \vspace{3pt}
 \centerline{\includegraphics[width=\textwidth]{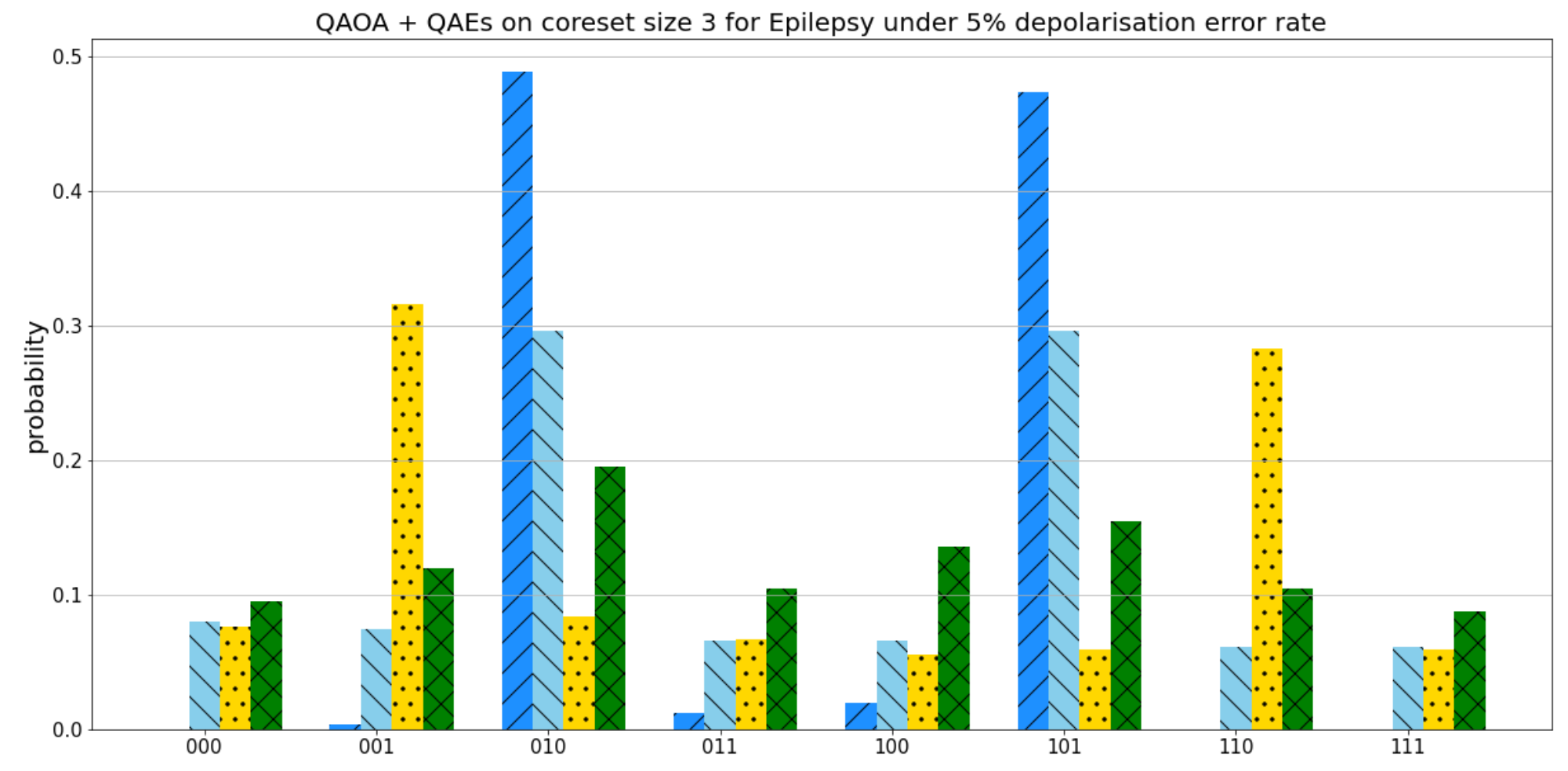}}
 \vspace{3pt}
 \end{minipage}
 
 \caption{QAEs result}
 \label{QAEs}
 \end{figure}

The quantum input to the experiment is a coreset of size 3 obtained from the Epilepsy data set. The QAOA circuit is run under the 2\% depolarisation error rate and 5\% depolarisation error rate respectively, and the bit-flip error is manually set after the QAOA circuit to simulate the fully collapsed situation. The QAEs is applied to it to correct the collapsed state. The results are shown in Figure \ref{QAEs}. The dodger blue columns shaded with "/" are the results obtained by the QAOA circuit on a noise-free quantum simulator. The $\ket{010} $ and $\ket{101}$ pairs have the highest probability of symmetric quantum states, and after introducing depolarisation noise, their probability drops significantly, and the decrease is more pronounced in a depolarisation error rate of 5\%, the results are shown in the sky blue with "\textbackslash" shading superior. In this situation, $\ket{010} $ and $\ket{101}$ pairs also have the highest probability relative to other quantum states. The worst-case scenario is a complete collapse of the quantum state, which could be caused by the bit-flip error, shown in the yellow column with a dot. 

QAEs can correct the collapsed state back for the QAOA circuit. It has been proven by Achache \textit{et al.} \cite{achache2020denoising} that the QAEs can do the quantum noise cancellation for the GHZ circuit and this experiment extended it to the QAOA circuit. By giving a pure circuit as target and a noisy circuit as input, QAEs learn parameters $K$ that can transform the error state back to the pure state. We trained the QAEs in the simulator using a pure QAOA circuit as the target and a noisy QAOA circuit as input. The noise comes from bit-flip error and the error rate is set to 20\%. We run 150 epochs, each epoch takes about 270 seconds. The effect of trained QAEs on the fully collapsed quantum state is shown in green columns with "x" shadowed. QAEs give the highest probability back to $\ket{010} $ and $\ket{101}$ whether at 2\% depolarisation error rate or 5\% depolarisation error rate, and the state at 2\% depolarisation error rate is closer to the ideal one. The limitation of QAEs is the training time. For the 5-qubit QAOA circuit and [5,1,5] structure of QAEs, 12 qubits are required, and the training time is around 2 hours per epoch, but the desired effect is not well achieved after 100 epochs. It requires a large amount of resources and time.

\section{CONCLUSIONS AND FUTURE DIRECTIONS}\label{conclusion}

In this work, we have analysed the impact of the coreset selection on the quantum K-Means clustering performance, including the analysis of influencing factors such as the coreset construction algorithm, the coreset size, and quantum noise. We found that both BFL16 and ONESHOT coreset construction algorithms perform well at summarizing robustly clustered data sets. The impact of these algorithms on the performance of quantum K-Means clustering is not remarkable, when using classical methods for segmentation and clustering operations, quantum K-Means clustering can achieve over 95\% accuracy on all 6 data sets investigated in this study. The impact of the coreset size is also an important parameter of interest. For most data sets, a larger coreset would reduce the performance of QAOA with the Nelder-Mead optimizer. This is because the number of quantum states brought by each additional qubit increases exponentially, making it difficult for QAOA and optimizers to find approximate solutions, and wrong segmentation and clustering operations directly affect accuracy. Meanwhile, a larger coreset means larger quantum circuits, which introduce more quantum noise. It turns out that QAEs can be extended to QAOA circuits with desirable results.

Some parts of our research can be expanded in a future study. A possible research direction is to find a way to make the coreset with limited size but with better summary of the data set. When using the coreset construction algorithm to obtain a coreset $M$ with size $m$, instead of using coreset $M$, smaller size of coreset $N$ with size $n (n << m)$ can be taken as the final coreset. Coreset $N$ should not be a subset of $M$, because $M$ already represents the distribution of the dataset, and $N$ being a subset of $M$ means that certain accuracy will be lost, which will bring more negative factors. Another possible direction of research could be around the performance of classical optimisers for QAOA circuit. More sophisticated optimisers such as Simultaneous Perturbation Stochastic Approximation (SPSA) optimizers or Adam and AMSGRAD optimizers can be studied which might allow better accuracy for larger QAOA circuits corresponding to bigger coreset sizes. Extending research to real quantum hardware is worth considering. We introduce quantum noise through a depolarizing noise model on the simulator. If on the real quantum hardware, the structure of the hardware, the relative error rate of each qubit, and other types of quantum errors can all affect the results. 

Overall, our study provides useful insights into the application of quantum algorithms for data science problems where the data size has been reduced by coreset selection. We show that this a promising area of further research which could enable near future quantum devices to handle data science applications of practical relevance.

\begin{acks}
This work was supported by IBM Q Hub at the University of Melbourne. Sarah Erfani is in part supported by Australian Research Council (ARC) Discovery Early Career Researcher Award (DECRA) DE220100680.
\end{acks}

\bibliographystyle{ACM-Reference-Format}
\bibliography{reference}

\end{document}